\documentclass[final,english]{elsarticle}
\usepackage[T1]{fontenc}
\usepackage[latin9]{inputenc}
\usepackage{varioref}
\usepackage{units}
\usepackage{amsmath}
\usepackage{amssymb}
\usepackage{graphicx}
\usepackage{esint}
\usepackage{caption}
\usepackage{subcaption}
\usepackage{appendix}

\pdfpageheight\paperheight
\pdfpagewidth\paperwidth

\journal{Elsevier}
\usepackage{upgreek}
\usepackage[bookmarks,bookmarksopen,bookmarksdepth=6]{hyperref}
\usepackage{cases}
\usepackage{url}
\usepackage{lineno}

\usepackage{a4wide}

\usepackage{newtxtext,newtxmath,amsmath}
\makeatother

\usepackage{babel}
\begin{document}


\title{Can the electric field in radiation-damaged silicon pad diodes be determined by admittance and current measurements?}

\author[]{R.~Klanner \corref{cor1}}
\author[]{J.~Schwandt}

\cortext[cor1]{Corresponding author. Email address: Robert.Klanner@desy.de,
 Tel.: +49 40 8998 2558}
\address{ Institute for Experimental Physics, University of Hamburg,
 \\Luruper Chaussee 149, 22761, Hamburg, Germany.}



\begin{abstract}

 A method is proposed for determining the electric field in highly-irradiated silicon pad diodes using admittance-frequency ($Y$-$f$), and current measurements ($I$).
 The method is applied to $Y$-$f$ and $I$ data from square $n^+p$ diodes of 25\,mm$^2$ area irradiated by 24 GeV/c protons to four 1 MeV neutron equivalent fluences between $3 \times 10^{15}$\,cm$^{-2}$ and $13 \times 10^{15}$\,cm$^{-2}$.
 The measurement conditions were:
 Reverse voltages between 1\,V and 1000\,V,
 frequencies between 100\,Hz and 2\,MHz and
 temperatures of $-20\,^{\,\circ}$C and  $-30\,^{\,\circ}$C.
 The position dependence of the electric field is parameterised by a linear dependence at the two sides of the diode, and a constant in the centre.
 The parameters as a function of voltage, temperature and irradiation fluence are determined by fits of the model to the data.
 For voltages below about 300\,V all data are well described by the model, and the results for the electric field agrees with expectations:
 Depleted high-field regions towards the two faces and a constant low electric field in the centre, with values which agrees with the field in an ohmic resistor with approximately the intrinsic resistivity of silicon.
 For conditions at which the low field region disappears and the diode is fully depleted, the method fails.
 This happens around 300\,V for the lowest irradiation fluence at $-30\,^{\,\circ}$C, and at higher voltages for higher fluences and lower temperatures.
 In the conclusions the successes and problems of the method are discussed.

 \end{abstract}


\maketitle
 \tableofcontents
 \pagenumbering{arabic}

\newpage

\section{Introduction}
 \label{sect:Introduction}

 The measurement of the capacitance-voltage, $C$-$V$ characteristics of diodes is a standard tool to determine the doping profiles in diodes and MOSFETs~\cite{Schroder:2006}, from which the electric field can be calculated.
 For pad diodes made of high-quality silicon it is found that for frequencies, \emph{f}, up to a few MHz, where the measurements are typically made, $C$ does not depend on $f$, and the doping profile as a function of the distance $x$ from the junction can be obtained from
 \begin{equation}\label{equ:CV}
   N_d(x) = \frac{2}{q_0 \cdot \varepsilon _{Si} \cdot  A^2 \cdot \mathrm{d}(1/C^2)/\mathrm{d}U }
 \end{equation}
  with the doping density $N_d(x)$ at the depletion depth $x(U) = \varepsilon _{Si} \cdot  A / C(U)$, with
  the elementary charge $q_0$,
  the dielectric constant of silicon $\varepsilon _{Si}$,
  the area of the diode $A$, and
  the reverse voltage $U$.
 This formula is valid, as long as the gradient of the doping profile $(1/N_d) \cdot \mathrm{d} N_d /\mathrm{d} x \ll L_D$, with the Debye length, $L_D$~\cite{Schroder:2006}, the transverse position dependence of $N_d$ can be neglected, the transverse dimensions of the diode are large compared to its thickness, and the measurement frequency is not too high\,\cite{Schwandt:2020}.

 For semiconductors with a high density of states in the band gap, a frequency dependence $C(f)$ is expected.
 Radiation damage of silicon by neutrons or high-energetic particles, which is addressed in this paper, produces such states.
 The frequency dependence of the capacitance for radiation-damaged silicon diodes is well established, and has already been observed in the 1960-ies~\cite{Wilson:1968}.
 In the following a selection of results from $Y$\,measurements of irradiated silicon diodes is presented.
 Table\,\ref{tab:Experiments} summarizes relevant parameters of these studies.

 \begin{table}[!ht]
  \centering
   \begin{tabular}{c|c|c|c|c|c|c|c}
 Type & initial $\rho $\,[$\Omega $\,cm] & Irrad. & $\Phi _\mathit{eq}$\,[cm$^{-2}$] & $U$ [V] & $f$ [Hz] & $T$ [K] & Ref. \\ \hline \hline
$n$ & unknown & $n$ & $10^{15}$ & $0 - 2.5$ & $500 - 500$\,k & 90--310 & \cite{Tokuda:1977} \\
$n$ & 10--10\,k & $n$ & $(7 - 800)\times 10^{10}$ & 3--100 & $ 500 - 1$\,M & RT & \cite{Li:1990} \\
$n$ & 5 k, 9.7 k & $n$ & $1.2 \times 10^{14}$ & $5 - 20$ & $0-150$\,k & 10--270& \cite{Croitoru:1997} \\ \hline
 $p$ & 3.4\,k & $p$ & $(3 - 13) \times 10^{15}$ & $0-1000$ & $100-2$\,M & 243, 253 & this work \\
   \end{tabular}
    \caption{Selected experiments studying the frequency dependence of irradiated silicon diodes. In column 1, $n$ and $p$, refer to $n$-type and $p$-type doping, respectively, and in column 3, $n$ to low energy (approximately 1\,MeV) neutron- and $p$ to 24\,GeV/c proton-irradiation. RT in column 7 stands for \emph{Room Temperature}, which is assumed, as the temperature is not given in the publiaction.
    \label{tab:Experiments} }
 \end{table}

 In Ref.~\cite{Tokuda:1977} $p^+n$~diodes have been irradiated with reactor neutrons to a fluence of $\Phi = 10^{15}$~cm$^{-2}$ and the admittance, $Y(U,f,T)$ measured for reverse voltages up to 2.5~V,  frequencies between 5 and 50~kHz and temperatures, $T$, between 90 and 310~K.
 Using the \emph{Admittance Spectroscopy} method\,\cite{Losee:1972, Losee:1975, Vincent:1975}, three defect levels in irradiated $n$-type silicon have been identified and their energy levels, electron cross sections and introduction rates determined.

 In Ref.~\cite{Li:1990} $p^+n$ diodes with resistivities between  10~$\Omega \cdot$cm and 10~k$\Omega \cdot$cm were irradiated by neutrons of energies between 10~keV and 2.2~MeV using the $^7$Li(p,n)~reaction up to a fluence of $\Phi = 7.8 \times 10^{12}$\,cm$^{-2}$.
 For voltages between 3 and 100~V, $C$-$V$~measurements with frequencies, $f$, between 500~Hz and 1~MHz were performed.
 A strong $f$~dependence of $C$ at high $\Phi $~values has been observed:
 At $\Phi = 7 \times 10^{10}$~cm$^{-2}$, the voltage dependence is $C \propto U^{-1/2}$ independent of $f$, which agrees with the results for the non-irradiated diode.
 At $\Phi = 7.8 \times 10^{12}$~cm$^{-2}$, for voltages below full depletion, $C$ decreases with increasing $f$, and for $f \gtrsim 100$~kHz, $C$ is equal to the geometric capacitance $C_{geom} = \varepsilon_{Si} \cdot A/d$ independent of voltage.
 In addition, the full depletion voltage of the 10~k$\Omega \cdot$cm diode decreases from 60~V to 20~V, and the $f$~dependence of $C$ is a function of the initial dopant concentration.
 All three effects can be described by a model which assumes Shockley-Reed-Hall statistics and two deep acceptor levels.
 The authors of Ref.~\cite{Li:1990} conclude:
 \emph{Considerable information about the trapping nature of radiation induced defects can be obtained from the frequency dependence of the $C(U)$ characteristics of junction diodes}.
 It should be noted that an extrapolation of the model of Ref.~\cite{Li:1990} to higher fluences fails in describing the performance of highly-irradiated silicon sensors.

 In Ref.~\cite{Croitoru:1997} $p^+n$ diodes with resistivities of 5 and 9.7~k$\Omega \cdot $cm were irradiated by neutrons to a fluence of $1.2 \times 10^{14}$~cm$^{-2}$ and the admittance $Y$ measured at reverse voltages of 5, 10 and 20~V, for frequencies between 0 and 150~kHz and temperatures between 10 and 270~K.
 At a fixed voltage, the measured admittance is fitted by the admittance of an equivalent electric circuit consisting of a capacitor
 $C_d = \varepsilon _{Si} \cdot A / w $, which is in series with a capacitor
 $C_s = \varepsilon _{Si} \cdot A / (d - w) $ and a resistor
 $R_s = \rho \cdot (d - w) / A$ in parallel;
 $C_d$ represents the capacitance of the depletion region of depth $w$, and $C_s$ and $R_s$ the capacitance and resistance of the non-depleted region with resistivity, $\rho $, respectively.
 For $T \gtrsim 50$~K, which approximately corresponds to the freeze-out temperature, the model provides a good description of the experimental data, and it is found that above a critical fluence of order $10^{14}$~cm$^{-2}$ the value of $\rho $ in the non-depleted region is significantly higher than before irradiation.
 The increase in $\rho $ is solely attributed to the increase in generation--recombination rate due to the radiation-induced states in the silicon band gap, which via the relation
 $n_h \cdot n_e = n_i^2$ results in a decrease of the density of free charge carriers;
 $n_i$, $n_h$ and $n_e$ are the intrinsic carrier density, and the density of holes and electrons, respectively.
 Recent TCAD simulations \cite{Schwandt:2020} have confirmed that for $\Phi \approx 10^{14}$~cm$^{-2}$, as a result of radiation-induced states in the silicon band gap, the intrinsic resistivity of silicon is reached, which is orders of magnitude higher than the resistivity of the silicon before irradiation.
 Whereas in the first papers discussed the frequency dependence of the charging and discharging of the radiation-induced states is considered to cause the $f$~dependence of the admittance, in the latter two publications it is ascribed to the radiation-induced change in resistivity of the non-depleted region.
 Which of these two effects gives the main contribution to the frequency and voltage dependence of the admittance in radiation-damaged silicon diodes remains an open question, which is addressed in this paper.

 In this paper admittance measurements are presented for $n^+p$ pad diodes irradiated by 24~GeV/c protons to 1~MeV neutron equivalent fluences, $\Phi _{eq}$, between 3 and $13 \times 10^{15}$~cm$^{-2}$ for reverse voltages up to 1000~V, frequencies between 100~Hz and 2~MHz at temperatures of $-30^{\,\circ } $C and $-20^{\,\circ } $C.
 In addition, the voltage and temperature dependence of the dark current $I(U,T)$ has been measured.
 For every $U$, $T$ and $\Phi _{eq}$, the $f$\,dependence of the admittance $Y$ is fitted by a model using a parametrisation of the position-dependent resistivity, from which, using the $I(U)$\,data, the position-dependent electric field in the diode is obtained.

 In the next section the pad diodes investigated and the measurements are presented.
 A discussion of the model used to describe the data and of the analysis method follows.
 The results of the analysis are presented in Sect.\,\ref{sect:Analysis}.
 In the conclusions the main results are summarized and the applicability of the method discussed.

  \section{Pad diodes investigated and measurements}
  \label{sect:Sensors}

 The pad diodes investigated have been produced by Hamamatsu Photonics K.K.~\cite{Hamamatsu} on $p$-type silicon with a boron doping of about $3.8 \times 10^{12}$~cm$^{-3}$.
 The area of $n^+$ implant is $A = 5$~mm $\times $ 5~mm.
 The $n^+$ implant is surrounded by a guard ring.
 The total thickness of the diodes, measured with a caliper, is $202~\upmu$m with an uncertainty of $2 ~\upmu$m, which, after subtracting the depth of $2\,\mu$m of the $n^+$ and $p^+$ implants, gives an active thickness of $d = 198 \pm 3~\upmu$m.

 The irradiations took place at the CERN PS~\cite{CERN_Irr}.
 The 1~MeV neutron equivalent fluences, $\Phi _{eq}$, were 3.0, 6.07, 7.75 and $13 \times 10^{15}$~cm$^{-2}$, where a hardness factor 0.62 has been used.
 The estimated uncertainty of $\Phi _{eq}$ is $ 10$~\%.
 After irradiation the diodes were annealed for 80 minutes at $60~^\circ$C and then stored in a freezer at temperatures below  $-20~^\circ$C.

 The current measurements were made on a probe station with a temperature-controlled chuck using a Keithley 6517A multimeter.
 The admittance measurements were performed on the same probe station with an Agilent 4980A LCR meter.
 Both pad and guard ring were connected to ground and the $p^+$~back implant was set to the reverse voltage $U$ using a Keithley 6517A multimeter.
 The AC-voltage used was 500~mV.
 The inverse of the parallel resistance, $1/\mathit{Rp}$, and the parallel capacitance, $\mathit{Cp}$, were recorded.

 $Y$- and $I$-data for the following conditions were recorded:\\
\begin{tabular}{ll}
  Irradiation: & $\Phi _{eq} = (3.0,\,6.07,\,7.75,\,13.0) \times 10^{15}\,\mathrm{cm}^{-2}$ \\
  Temperature: & $ T = (-20,\,-30)\, ^\circ $C \\
  Voltage:     & $U = 1\, \mathrm{to} \, 1000$\,V (125 values) \\
  Frequency:   & $f = 100$\,Hz to 2 MHz (16 values) \\
\end{tabular}

 The data at $f = 1.5$ and 2.0\,MHz are not used in the analysis, as the decoupling circuit caused an increase of the measured $\mathit{Cp}$ at these frequencies.

 \section{Admittance model and analysis method}
  \label{sect:Model}

\begin{figure}[!ht]
   \centering
    \includegraphics[width=0.25\textwidth]{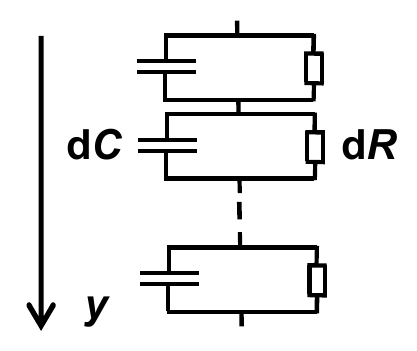}
   \caption{Electrical model of the pad diode:
    $\mathrm{d}C(y)$ and $\mathrm{d}R(y)$ are the differential capacitance and differential resistance of the sensor element between $y$ and $y + \mathrm{d}y$.
    The impedance of the entire sensor is obtained by integrating
    $\mathrm{d}Z(y) = \big(1/\mathrm{d}R(y) + i \, \omega \, \mathrm{d}C(y) \big) ^{-1}$ over the depth $d$ of the diode. }
  \label{fig:Model}
 \end{figure}

 Fig.~\ref{fig:Model} shows the electrical model used to describe the impedance measurements.

 The diode is divided into slices of depth d$y$, each one described by the resistor
 d$R = \rho (y) \cdot \mathrm{d}y/A$ in parallel with the capacitor
 d$C = \varepsilon _{Si} \cdot A/\mathrm{d}y$.
 The total complex resistance, $Z(\omega )$, is obtained by summing the complex series resistors of the individual slices
 \begin{equation}\label{equ:Ztot}
   Z(\omega) = \int _0 ^d \Big( \frac{1}{\mathrm{d}R} + i \cdot \omega \cdot \mathrm{d}C \Big) ^{-1}
   = \frac{1}{A} \cdot \int _0 ^d \frac{\rho (y) \cdot \mathrm{d}y}{1 + i \cdot \omega \cdot \varepsilon _{Si} \cdot \rho (y)}
   = \frac{1}{A} \cdot \int _0 ^d \frac{\mathrm{d}y }{1/\rho (y) + i \cdot \omega \cdot \varepsilon _{Si}},
 \end{equation}
 and $Y(\omega ) = 1/\mathit{Rp} + i \cdot \omega \cdot \mathit{Cp} = Z(\omega ) ^{-1}$.
 The model assumes that the frequency dependence, $Y(\omega)$, is solely caused by the resistivity without any contribution from the charging and discharging of radiation-induced traps.
 In addition, it assumes that the active thickness of the diode, $d$, and dielectric constant, $\varepsilon _{Si}$, do not change with fluence $\Phi $.
 The latter two assumptions could be checked with the data:
 At high frequencies, where according to Ref.\,\cite{Schwandt:2020} the geometric capacitance of the diode is expected, the measured capacitance is independent of $\Phi $ within the experimental uncertainties of about $\pm 0.2$\,\%.

 Inspection of Eq.~\ref{equ:Ztot} shows \footnote{The value of $1 / (\omega \cdot \varepsilon _{Si}) \approx 1.6 \times 10^8 ~ \Omega \cdot $cm at $f = 1$~kHz. }:
 \begin{itemize}
  \item For $\rho(y) \gg 1 / (\omega \cdot \varepsilon _{Si}) $ in the entire diode, which corresponds to a fully depleted, non-irradiated diode, $Z(\omega) =  1 / (i \cdot \omega \cdot C_\mathit{geom} ) $ with the geometrical capacitance
      $ C_\mathit{geom} =\varepsilon _{Si} \cdot A / d $.
  \item For $\rho(y) \gg 1 / (\omega \cdot \varepsilon _{Si}) $ up to a depth $w$ and $\rho = 0$ for the remainder, which correspond to a partially depleted non-irradiated diode, $Z(\omega) = 1/ (i \cdot \omega \cdot C_w)$ with $C_w = \varepsilon _{Si} \cdot A / w $, the capacitance of the depletion depth $w$.
  \item for $\rho(y) \ll 1 / (\omega \cdot \varepsilon _{Si}) $ in the entire diode, which corresponds to a bias close to 0\,V, $Z(\omega ) = \rho \cdot d / A$, the resistance of an ohmic resistor with the shape of the pad diode and the resistivity $\rho$.
 \end{itemize}

 The resistivity is related to the density of free electrons, $n_e$, and holes, $n_h$, and their mobilities $\mu _e$ and $\mu _h$ by
  $\rho(y) = \big( q_0 \cdot (n_e \cdot \mu _e + n_h \cdot \mu _h) \big) ^{-1} $.
 Ignoring the diffusion current, the relation between the resistivity, the electric field, $E$, and the current density, $j = I / A$, is given by Ohm's law
 \begin{equation}\label{equ:rho-E}
   \rho(y) = E(y)/j.
 \end{equation}
 In Appendix\,\ref{sect:Appendix} it is shown that using this relation results in systematically lower $E$\,values.
 However, the correction is small and can be neglected.
 In steady-state conditions $j$ does not depend on $y$, and $E(y)$ can be determined from $\rho (y)$ if $j$ is known.
 In addition, the relation $U = \int _0 ^d E(y) \cdot \mathrm{d}y$ gives the following constraint for $\rho(y)$:
 \begin{equation}\label{equ:Rconstraint}
   R = \frac{U}{I} = \frac{\int _0 ^d \rho(y) \cdot \mathrm{d}y}{A}.
 \end{equation}


 As $Z(\omega )$ is given by the integral of a function of $\rho (y)$ (Eq.\,\ref{equ:Ztot}), additional assumptions on the functional dependence of $\rho (y)$ are required for its determination.
 This problem exists already for non-irradiated pad diodes.
  In deriving Eq.\,\ref{equ:CV} it is assumed that the $p$-type sensor depletes from the $n^+p$ junction, that the space-charge density in the depletion region does not change with voltage, and that the electric field caused by the diffusion of holes at the $p\,p^+$\,junction can be ignored.
 Whereas these assumptions are essentially fulfilled for non-irradiated pad diodes, most of them are not valid after irradiation  with hadron fluences exceeding approximately $ 10^{13}$\,cm$^{-2}$:
  The resistivity and the effective doping at a given $y$\,value depend on the occupancy of the radiation-induced states in the band gap which is influenced by the position- and voltage-dependent densities of electrons and holes from the dark current.
 Even the sign of the effective doping can differ in different regions of the diode, and high-field regions, the so called \emph{double junction}~\cite{Eremin:1995}, are observed at the two surfaces of the diode.

 It is concluded that determining $\rho (y)$ using the integral equation Eq.\,\ref{equ:Ztot} with the constraint of Eq.\,\ref{equ:Rconstraint}, is an ill-conditioned mathematical problem without a unique solution.
 In this paper the problem is tackled by parameterising $\rho(y)$ guided by the results of Ref.~\cite{Eremin:1995, Klanner:2020}, where it is shown that
 three regions can be distinguished in partially-depleted irradiated pad diodes (the signs refer to a $p$-type sensor with $ y = 0$ at the $n^+p$ junction):
 \begin{enumerate}
   \item A high-field region close to the $n^+p$ junction with an electric field which decreases with increasing $y$, corresponding to a negative space charge.
   \item A low field region with a constant field, corresponding to zero space-charge density.
   \item A high-field region close to the $p\,p^+$ junction with an electric field which increases with increasing $y$, corresponding to a positive space charge.
 \end{enumerate}

 Regions 1 and 3 are the quasi-depleted, and region 2 the non-depleted regions.
 Above a certain voltage, which increases with irradiation and also depends on $T$, region~2 disappears.
  The sensor can be considered to be fully depleted, however, with a significant density of free charge carriers which are generated by the radiation-induced states and swept away by the electric field.

 The difference in the sign of the effective space-charge in regions 1 and 3 is caused by charge carriers trapped in the radiation-induced states\,\cite{Eremin:1995}:
 The electric field for reverse bias points in the $+ y$-direction, and holes drift in the $+ y$ and electrons in the $- y$\,direction.
 Therefore the current in region 1 is predominantly from electrons, and in region 3 from holes.
 The resulting difference of $n_e (y) $ and $n_h (y) $ results in different occupancies of the radiation-induced acceptor and donor states and thus in a position dependent effective space charge.

 Region 2, where generation and recombination are in equilibrium, has the properties of an ohmic resistor with intrinsic resistivity
 \begin{equation}\label{equ:rhoi}
   \rho _\mathit{intr} =\Big(2 \cdot q_0 \cdot n_i \cdot \sqrt{\mu _{e,0} \cdot \mu _{h,0} }\Big) ^{-1}.
 \end{equation}
 The intrinsic carrier concentration, which has an exponential $T$\,dependence, is denoted $n_i$, $q_0$ is the elementary charge, and $\mu _{e,0}$ and $\mu _{h,0}$ are the low-field mobilities of electrons and holes, respectively.
 For $T = - 20 \, ^\circ$C, $\rho _\mathit{intr} \approx  2.3 \times 10^7 \, \Omega \cdot$cm, which is orders of magnitude larger than the resistivity of a few k$\Omega \cdot$cm used for sensor fabrication.
 The equation for $\rho _\mathit{intr} $ follows from the mass-action relation $n_e \cdot n_h = n_i^2$ and from $n_e \cdot \mu _{e,0} = n_h \cdot \mu _{h,0}$ for equal generation and recombination of electrons-hole pairs.

 Based on these considerations and in particular on the results of Ref.\,\cite{Klanner:2020}, the following parametrisation of $\rho (y)$ is assumed:
 \begin{equation}\label{equ:rho_y}
   \rho (y) = \sqrt{\rho _i ^2 + \Big(\rho _\mathit{front} \cdot \max(0\,,1 - y/\lambda _\mathit{front}) \Big)^2 +
   \Big(\rho _\mathit{rear} \cdot \max(0\,,1 - (d - y)/\lambda _\mathit{rear}) \Big)^2 },
 \end{equation}
 with \emph{front} referring to region 1, \emph{rear} to region 3, and $\rho _i$ the constant resistivity in region 2.
  Below full depletion, $\rho _i \approx \rho _\mathit{intr}$ is expected.
 As both $\rho _\mathit{front}$ and $\rho _\mathit{rear}$ are found to be large compared to $\rho _i$, $\rho _\mathit{front}$  and  $\rho _\mathit{rear}$  are approximately the resistivities at $y = 0$ and $y = d$, respectively, and $\lambda _\mathit{front} $ and $\lambda _\mathit{rear} $ are the widths of regions 1 and 3.
  The square root of the sum of squares is chosen to have smooth derivatives of $\rho(y)$ at the boundaries between the three regions.
 According to Eq.\,\ref{equ:rho-E}, $E(y) = j \cdot \rho(y)$, which implies that $\lambda _\mathit{front}$ and $\lambda _\mathit{rear}$ are also approximately the widths of the high-field regions adjacent to the $n^+p$\,junction and to the $p\,p^+$\,junction, respectively.
 It is also noted that a linear dependence of the electric field corresponds to a constant effective doping density.


 For a data set $(U$, $\Phi_\mathit{eq}$ and $T)$, the five parameters of Eq.\,\ref{equ:rho_y}, $\rho _i$, $\rho _\mathit{front}$, $\lambda _\mathit{front}$, $\rho _\mathit{rear}$, and $\lambda _\mathit{rear}$, are determined by minimising
  \begin{equation}\label{equ:Chitot}
   \chi ^2 = w_{\varphi} \cdot \chi _{\varphi} ^{\,2} + w_\mathit{Cp} \cdot \chi _\mathit{Cp} ^{\,2} + w_{I} \cdot \chi _{I} ^{\,2},
 \end{equation}
 with the three terms
 \begin{equation}\label{equ:Chisq}
   \chi_\varphi ^2 = \sum _f \bigg(\varphi _f ^{\,\mathit{meas}} - \varphi _f ^{{\,\mathit{model}}}\bigg)^2,
   \hspace{3mm}
   \chi_\mathit{Cp} ^2 = \sum _f\bigg(1-\frac{\mathit{Cp} _f ^{\,\mathit{meas}}} {\mathit{Cp} _f^{\,\mathit{model}}} \bigg)^2
   \hspace{2mm} \mathrm{and} \hspace{3mm}
   \chi _I ^2 = \bigg(1 - \frac{I} {U \cdot A} \int _0 ^d \rho(y)\,\mathrm{d} y \bigg)^2.
 \end{equation}

  In the fit $\mathit{Cp}$ and the phase  $\varphi = \arctan (\omega \cdot \mathit{Cp} \cdot \mathit{Rp} )$ are used.
 The reason for using $\varphi $ and not $\mathit{Rp}$ is that the uncertainty of $\varphi $ is approximately independent of $\varphi $, whereas the uncertainty of $\mathit{Rp}$ increases drastically when $\varphi $ approaches $90\,^\circ$;
 $\mathit{Rp}$ can even jump from $ + \infty $ to $ -  \infty$ because of measurement uncertainties.
 The frequencies selected for the fits are denoted \emph{f}, the measured parallel capacitances and phases $\mathit{Cp} _f^{\,\mathit{meas}}$ and $\varphi _f^{\,\mathit{meas}}$, and the model capacitances $\mathit{Cp}\, _f^{\,\mathit{model}}$, and phases $\varphi _f^{\,\mathit{model}}$, are obtained using Eqs.\,\ref{equ:Ztot} and \ref{equ:rho_y}.
 $I$ is the current at the reverse voltage $U$, and $A$ the area of the pad diode.
  The $w_i$\,values used are $w_\mathit{Cp} = w_\varphi =1$ and $w_I = 10^{-2}$.
 With this choice the contributions of the three terms to the $\chi ^2$ are similar.
  Fits with different $w$\,values were made as systematic check.
 As the admittance is not sensitive to the $y$\,position of a given $\rho$, interchanging $(\rho_\mathit{front},\,\lambda_\mathit{front})$ and $(\rho_\mathit{rear},\,\lambda_\mathit{rear})$ in Eq.\,\ref{equ:rho_y} gives exactly the same $\chi ^2$ values.
 To prevent a swapping between \emph{front} and \emph{rear} to occur, the condition $\rho_\mathit{front} > \rho_\mathit{rear} $ is imposed in the fit.

 It is noted that the parametrisation of Eq.\,\ref{equ:rho_y} is adequate for irradiated pad diodes below full depletion, where $\rho(y)$ varies by typically two orders of magnitude.
  For voltages above full depletion the variation of $\rho(y)$ is significantly less and the parametrisation is no more adequate.
 Although a description of the $\mathit{Cp} (f)$ and $\varphi (f)$ measurements can be achieved also in this case, strong correlations between the parameters are observed and the results for $\rho (y)$ cannot be trusted.
 For this reason, the emphasis will be on the analysis and results below full depletion.

 \section{Data analysis and results}
  \label{sect:Analysis}

 \subsection{Electric field below full depletion and intrinsic resistivity}
  \label{sect:rho_i}

 In this section the data for the irradiated pad diodes at voltages below 300\,V are presented and analysed, for which an ohmic low-field region with intrinsic resistivity, $\rho _\mathit{intr}$, in the central region is observed.
 The aim of the analysis is to check if the model discussed in Sect.\,\ref{sect:Model} is able to describe the frequency dependence of the admittance, $Y(f)$, and the current, $I$, for the different voltages, temperatures and irradiation fluences.
 It is also investigated if the values obtained for the low-field resistivity, $\rho _i$, agree with expectations, and if the $y$\,dependencies of the electric field, $E(y)$, is compatible with previous determinations at similar radiation fluences  and temperatures.
 The results also help to better understand the model assumptions.

\begin{figure}[!ht]
   \centering
   \begin{subfigure}[a]{0.5\textwidth}
    \includegraphics[width=\textwidth]{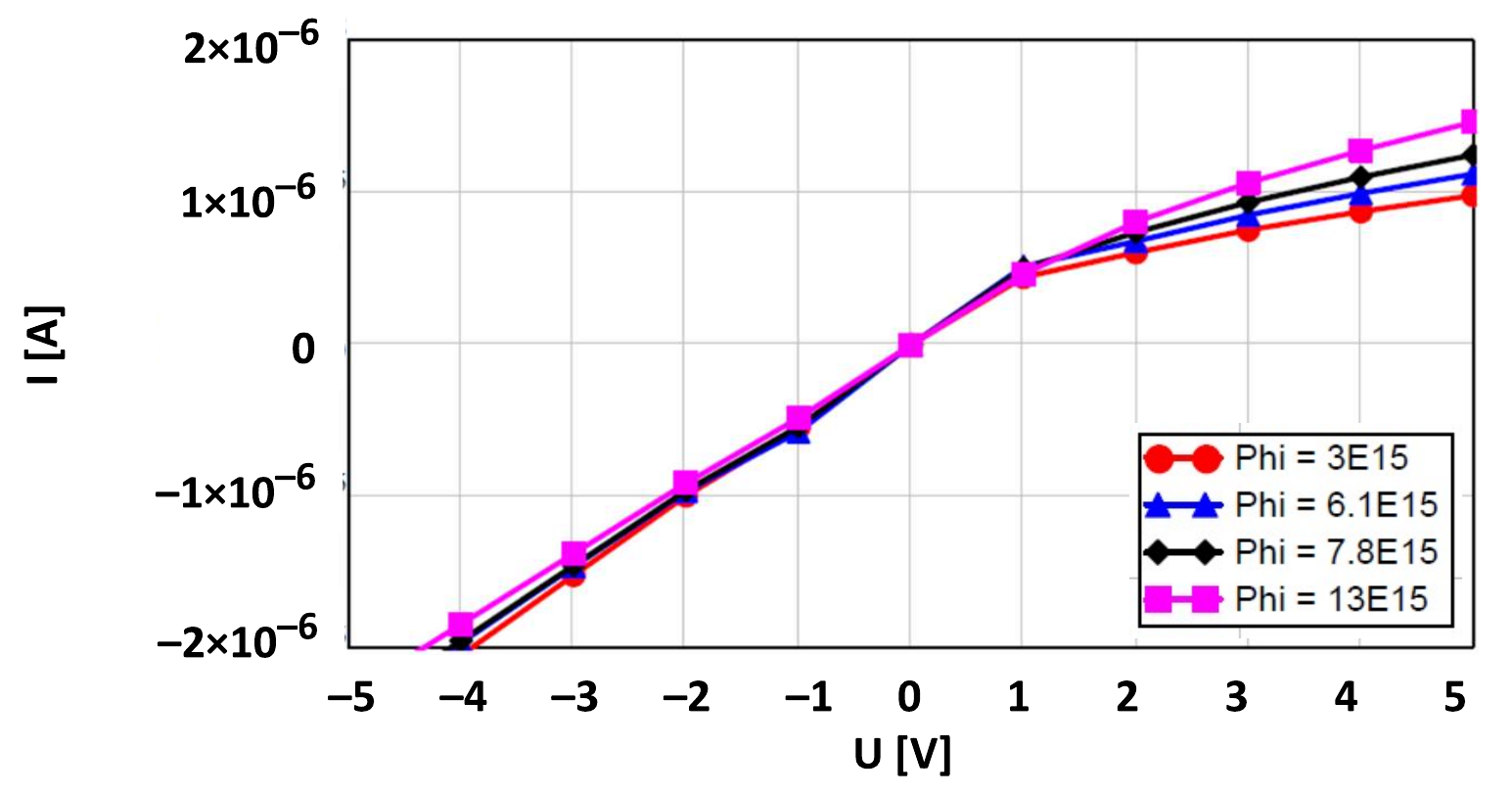}
    \caption{ }
    \label{fig:Fig_IVm20}
   \end{subfigure}%
    ~
   \begin{subfigure}[a]{0.5\textwidth}
    \includegraphics[width=\textwidth]{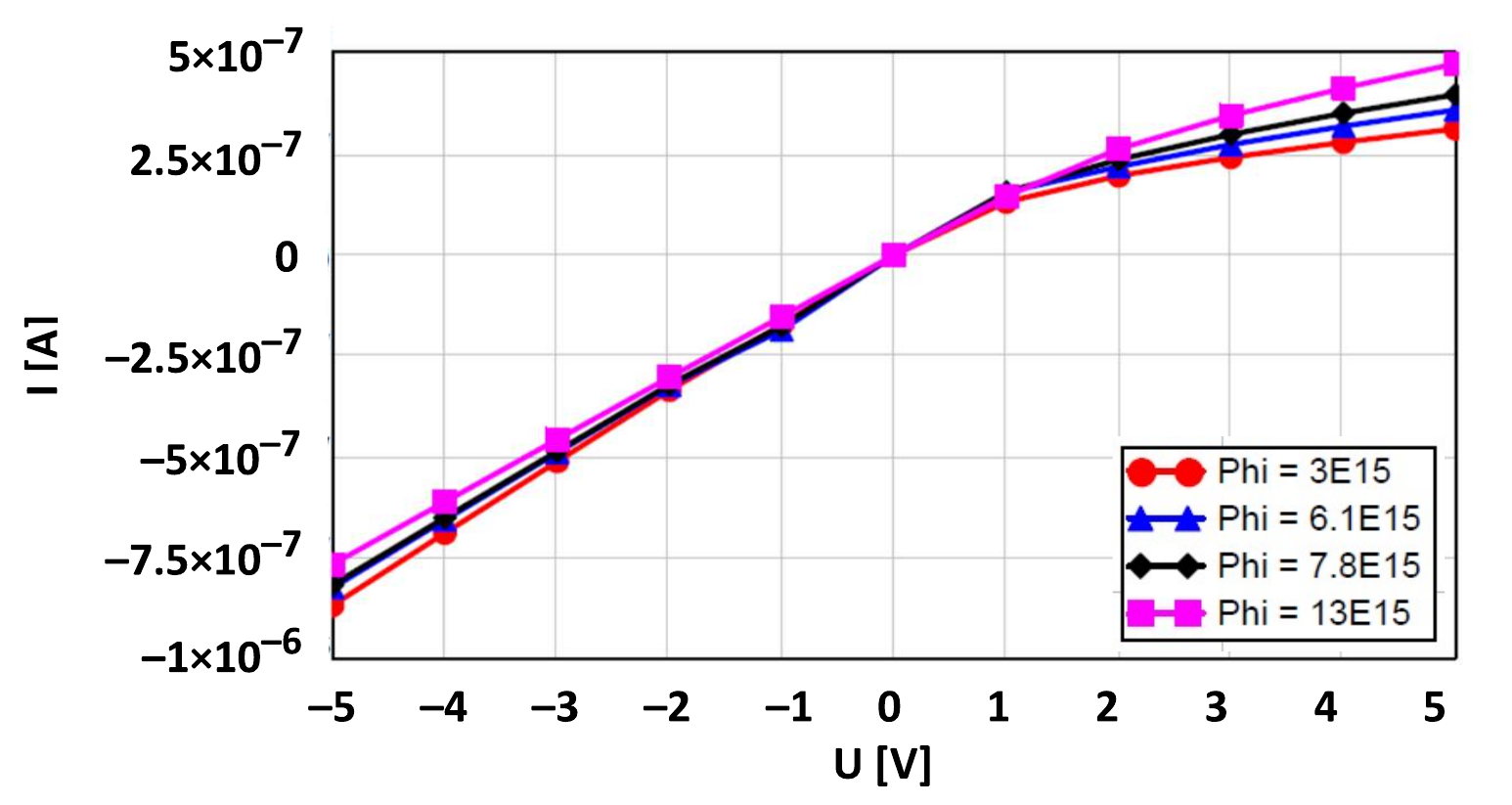}
    \caption{ }
    \label{fig:Fig_IVm30}
   \end{subfigure}%
   \caption{ $I-V$ measurement results for low forward ($-$) and low reverse $(+)$ voltages at (a) $T = - 20\,^\circ$C and (b) $T = - 30\,^\circ$C. }
  \label{fig:Fig_IV}
 \end{figure}

 Fig.\,\ref{fig:Fig_IV} shows the measured currents $I(U)$ for forward and reverse voltages up to $|U| = 5$\,V at $T = -20^{,\circ}$C and $- 30^{\,\circ}$C for the four irradiation fluences, $\Phi _{eq}$.
 For forward bias $I(U)$ is linear, like an ohmic resistor, with a resistance which is independent of $\Phi _{eq}$.
 The same linear dependence is observed for the reverse bias voltage up to $U \approx 1$\,V, as discussed in Ref.\,\cite{Scharf:2018}, where detailed $I(U)$\,data for $|U| < 1$ are presented.
 For higher reverse voltages $I(U)$ flattens towards a $U^{\approx \,0.6}$-dependence (Ref.\,\cite{Scharf:2018}).
 The resistivities for the linear region, $\rho_\mathit{IU}$, as well as the calculated intrinsic resistivity, $\rho _\mathit{intr}$ are given in Table\,\ref{tab:rho0}.
 For forward voltages the values of $\rho_\mathit{IU}$ and $\rho _\mathit{intr}$ essentially agree, whereas at the reverse voltage of 1\,V, $\rho _\mathit{IU}$ is 10 to 20\,\% higher than $\rho _\mathit{intr}$.
 The explanation of these observations is that the current in highly-irradiated sensors at low voltages is limited by the resistivity of the non-depleted bulk and not by the generation rate in the non-depleted regions.

 \begin{table}[!ht]
  \centering
   \begin{tabular}{c|c|c|c|c|c}
      $\Phi_{eq}$ [cm$^{-2}$]  & $3.0 \times 10^{15}$ & $6.1 \times 10^{15}$ & $7.8 \times 10^{15}$ & $13 \times 10^{15}$ & $\rho _\mathit{intr}$ \\
    \hline \hline
      $\rho _i (- 20^\circ $C) [$\Omega$ cm] & $26 \times 10^6$ & $26\times 10^6$  & $26\times 10^6$  & $28\times 10^6$ &  \\ \cline{1-5}
      $\rho _{IU_\mathit{rev}} (- 20^\circ $C)  [$\Omega$ cm]& $28 \times 10^6$ & $24\times 10^6$  & $25\times 10^6$  & $26\times 10^6$ &  $23\times 10^6$ \\
      $\rho _{IU_\mathit{forw}} (- 20^\circ $C)  [$\Omega$ cm]& $22 \times 10^6$ & $22\times 10^6$  & $23\times 10^6$  & $26\times 10^6$ &  \\
    \hline
      $\rho _i (- 30^\circ $C)  [$\Omega$ cm]& $78 \times 10^6$ & $82\times 10^6$  & $81\times 10^6$  & $83\times 10^6$ &  \\ \cline{1-5}
      $\rho _{IU_\mathit{rev}} (- 30^\circ $C)  [$\Omega$ cm]& $93 \times 10^6$ & $79\times 10^6$  & $79\times 10^6$  & $85\times 10^6$ &  $69\times 10^6$  \\
      $\rho _{IU_\mathit{forw}} (- 30^\circ $C)  [$\Omega$ cm]& $72 \times 10^6$ & $70\times 10^6$  & $70\times 10^6$  & $81\times 10^6$ &  \\
   \end{tabular}
    \caption{Comparison of $\rho _i $ in units of $\Omega \cdot $cm from the fits to the $ Y-f$~data at $U = 1 $~V with the resistivity, $\rho _\mathit{IU}$, from the reverse and forward $I-V$ measurements for $|U|$ between 0 and 1\,V, and the intrinsic resistivity, $\rho _\mathit{intr}$, calculated using Eq.\,\ref{equ:rhoi}.
    For the mobilities Ref.~\cite{Lombardi:1988} and for the intrinsic carrier concentration Ref.~\cite{Green:1990} have been used.
    The uncertainties of the determination of $\rho _i$ are dominated by systematic effects.
    Comparing the results of fits with different parametrizations of $\rho(y)$ and different $f$-intervals, errors of 10 to 20\,\% are estimated.
    The uncertainties for $\rho _{IU_\mathit{rev}}$ are similar, and for $\rho _{IU_\mathit{forw}}$ about a factor 2 smaller.
    \label{tab:rho0} }
 \end{table}

\begin{figure}[!ht]
   \centering
    \includegraphics[width=\textwidth]{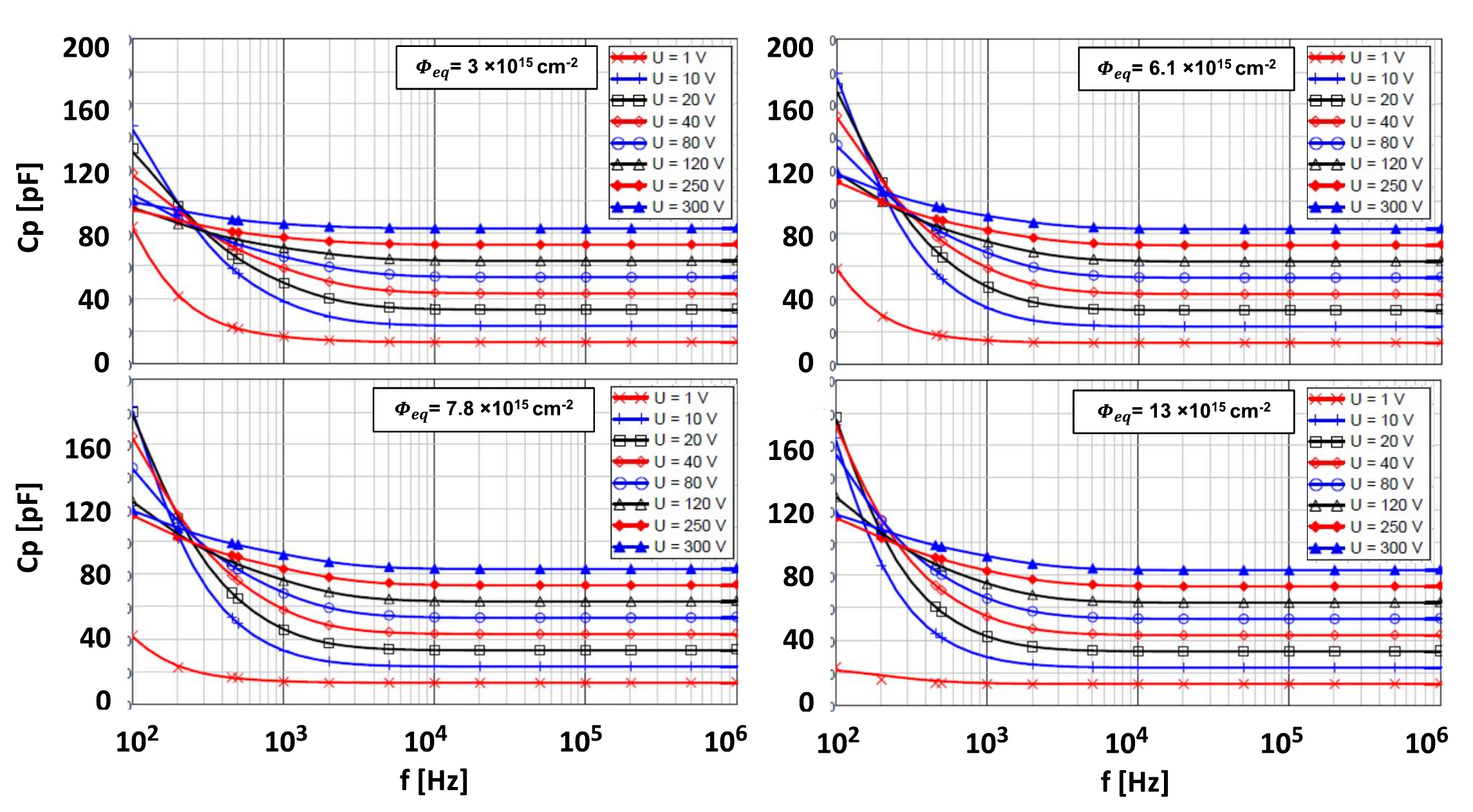}
   \caption{Comparison of the measured $\mathit{Cp}(f)$\,values (symbols) to the fit results (lines) at $T = - 20^{\,\circ}$C for voltages $U$ between 1\,V and 300\,V for the four $\Phi _\mathit{eq}$\,values.
   For clarity of presentation, the curves are shifted by 10\,pF for increasing $U$\,values. }
  \label{fig:Cp-m20}
 \end{figure}

\begin{figure}[!ht]
   \centering
    \includegraphics[width=\textwidth]{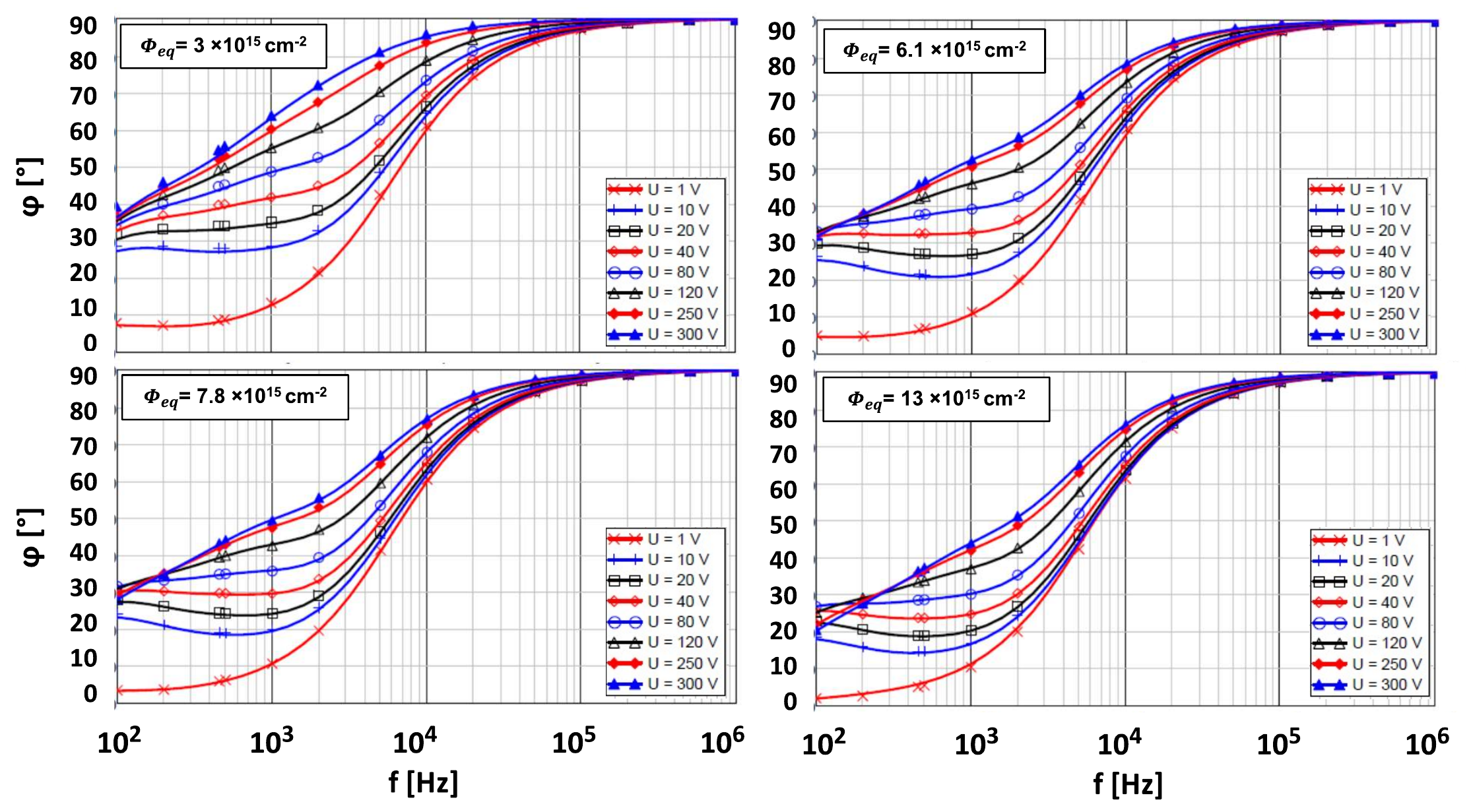}
   \caption{Comparison of the measured $\varphi(f)$\,values (symbols) to the fit results (lines) at $T = - 20^{\,\circ}$C for voltages $U$ between 1\,V and 300\,V for the four $\Phi _\mathit{eq}$\,values. }
  \label{fig:phi-m20}
 \end{figure}

\begin{figure}[!ht]
   \centering
    \includegraphics[width=\textwidth]{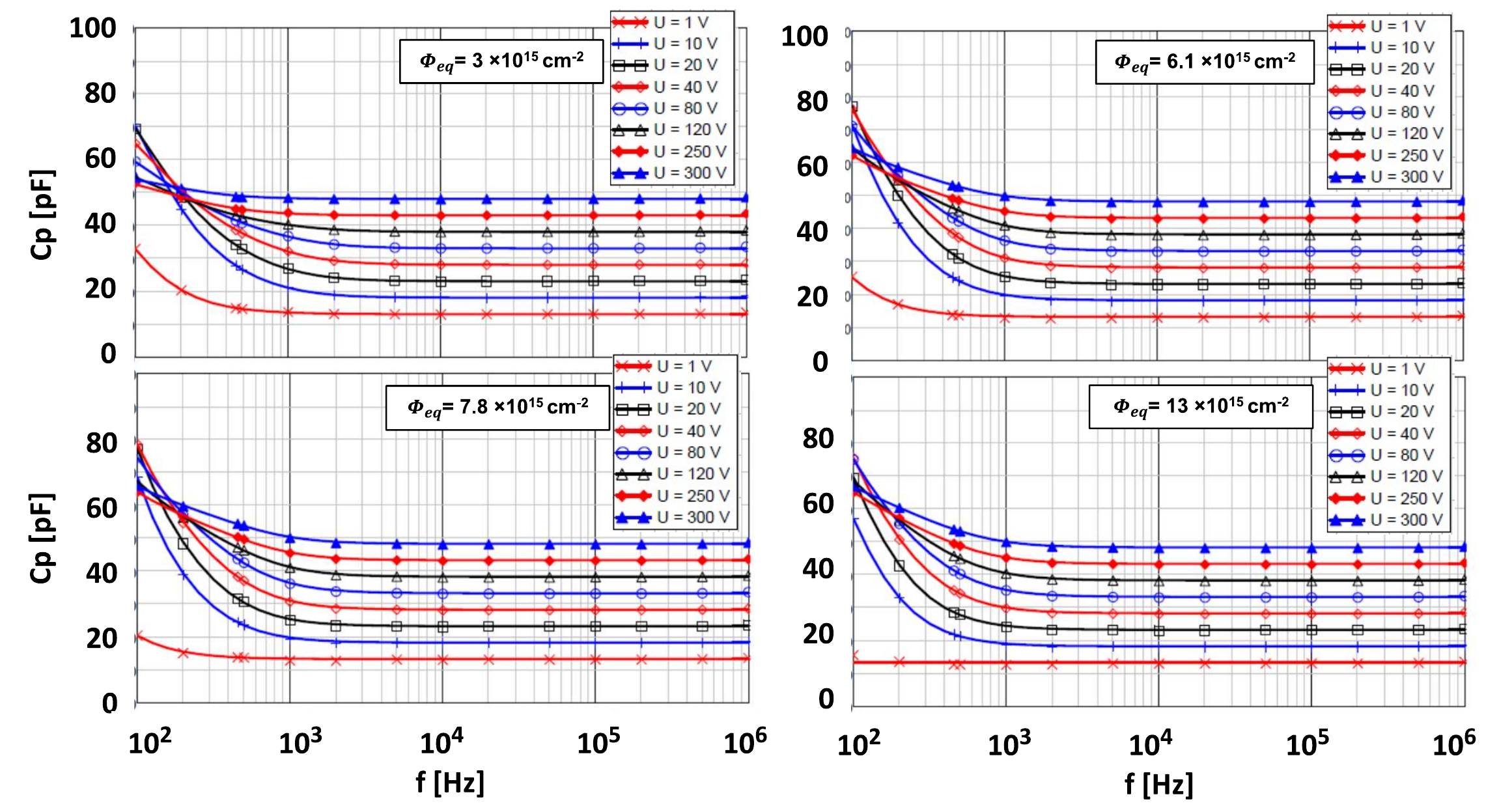}
   \caption{Comparison of the measured $\mathit{Cp}(f)$\,values (symbols) to the fit results (lines) at $T = - 30^{\,\circ}$C for voltages $U$ between 1\,V and 300\,V for the four $\Phi _\mathit{eq}$\,values.
   For clarity of presentation, the curves are shifted by 5\,pF for increasing $U$\,values. }
  \label{fig:Cp-m30}
 \end{figure}

\begin{figure}[!ht]
   \centering
    \includegraphics[width=\textwidth]{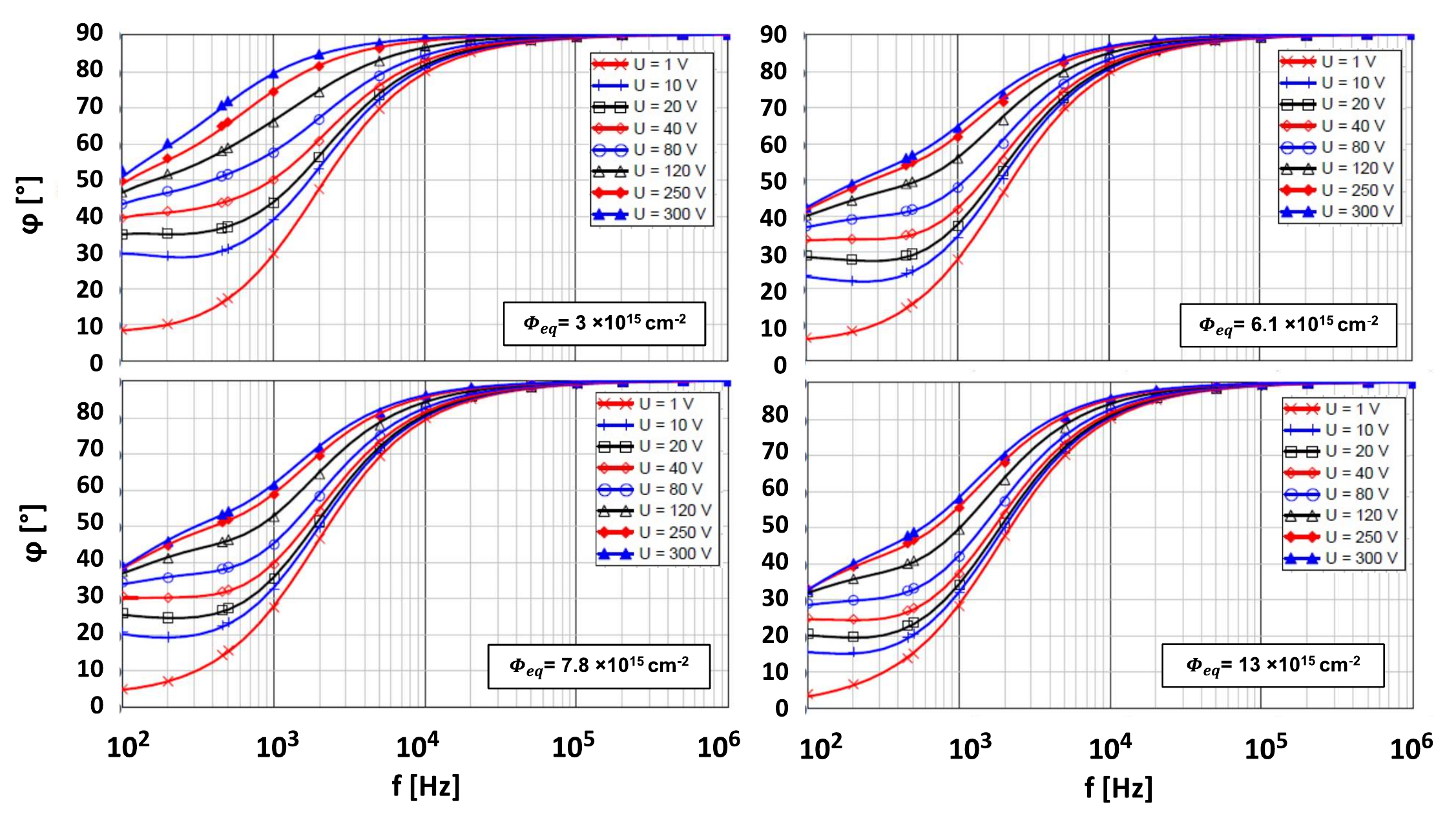}
   \caption{Comparison of the measured $\varphi(f)$\,values (symbols) to the fit results (lines) at $T = - 30^{\,\circ}$C for voltages $U$ between 1\,V and 300\,V for the four $\Phi _\mathit{eq}$\,values. }
  \label{fig:phi-m30}
 \end{figure}

 Next, the fits of the model to the admittance data are discussed.
  A selection of $\mathit{Cp}(f)$ and $\varphi (f)$ data for $U$ between 1\,V and 300\,V, the four $\Phi _\mathit{eq}$\,values and $T = -20^{\,\circ}$C and $-30^{\,\circ}$C are shown in Figs.\,\ref{fig:Cp-m20}, \ref{fig:phi-m20}, \ref{fig:Cp-m30}, and \ref{fig:phi-m30}, and compared to the results of the fits described in Sect.\,\ref{sect:Model}.
 It is remarkable how well the quite complex dependencies are described by the model:
  Typical deviations are 1\,\% for $\mathit{Cp}$ and $1^{\,\circ}$ for $\varphi $.
 It is noted that the frequency-dependent charging and discharging of the radiation-induced states in the band gap, as discussed in Refs.\,\cite{Tokuda:1977, Li:1990}, is not required to describe accurately the experimental data.

\begin{figure}[!ht]
   \centering
    \includegraphics[width=\textwidth]{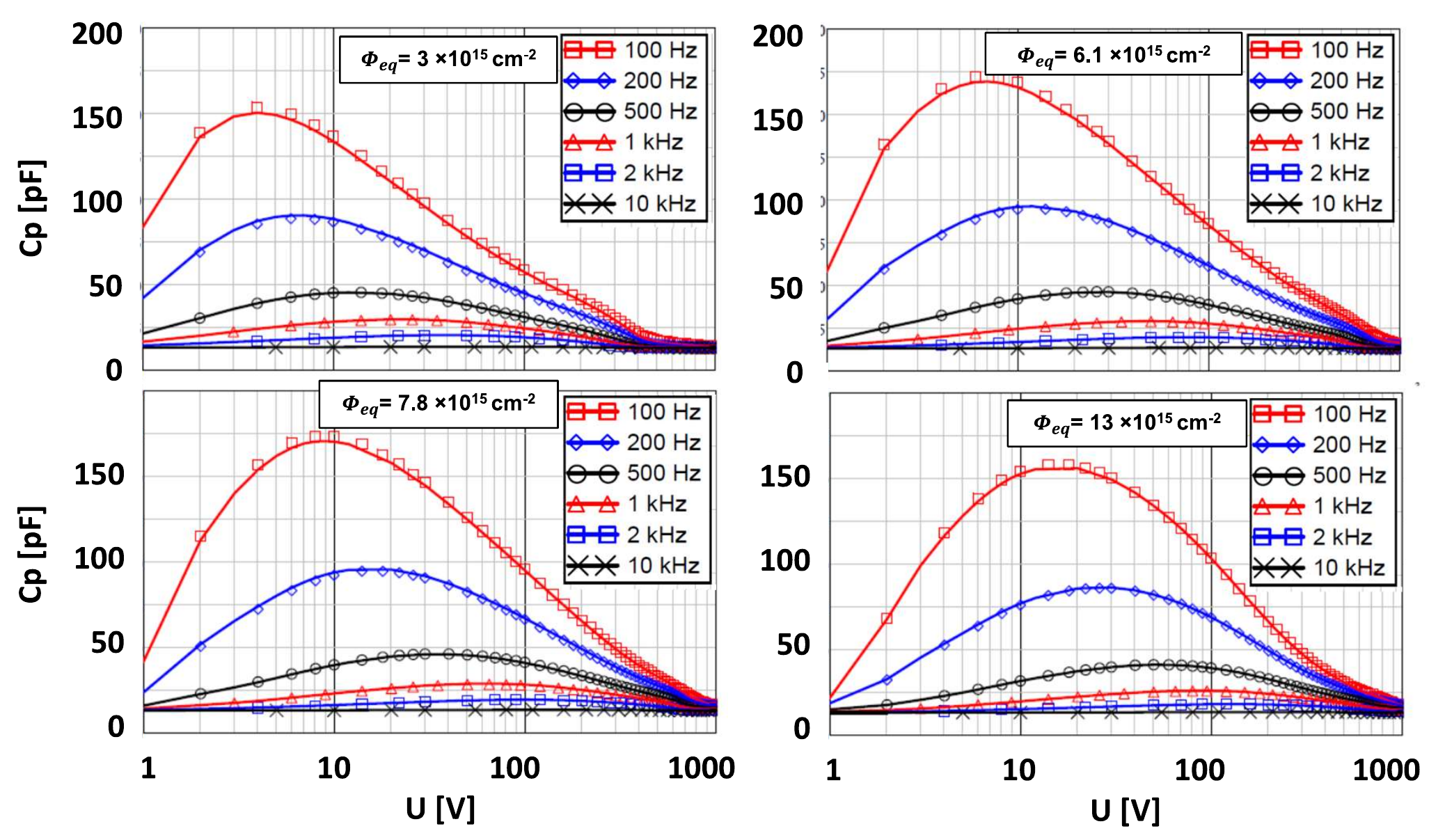}
    \caption{Voltage dependence of $\mathit{Cp}$ for selected frequencies at $T = - 20^{\,\circ}$C.
     The symbols are the experimental data, and the lines the results of the fits. }
  \label{fig:CpU-m20}
 \end{figure}

\begin{figure}[!ht]
   \centering
    \includegraphics[width=\textwidth]{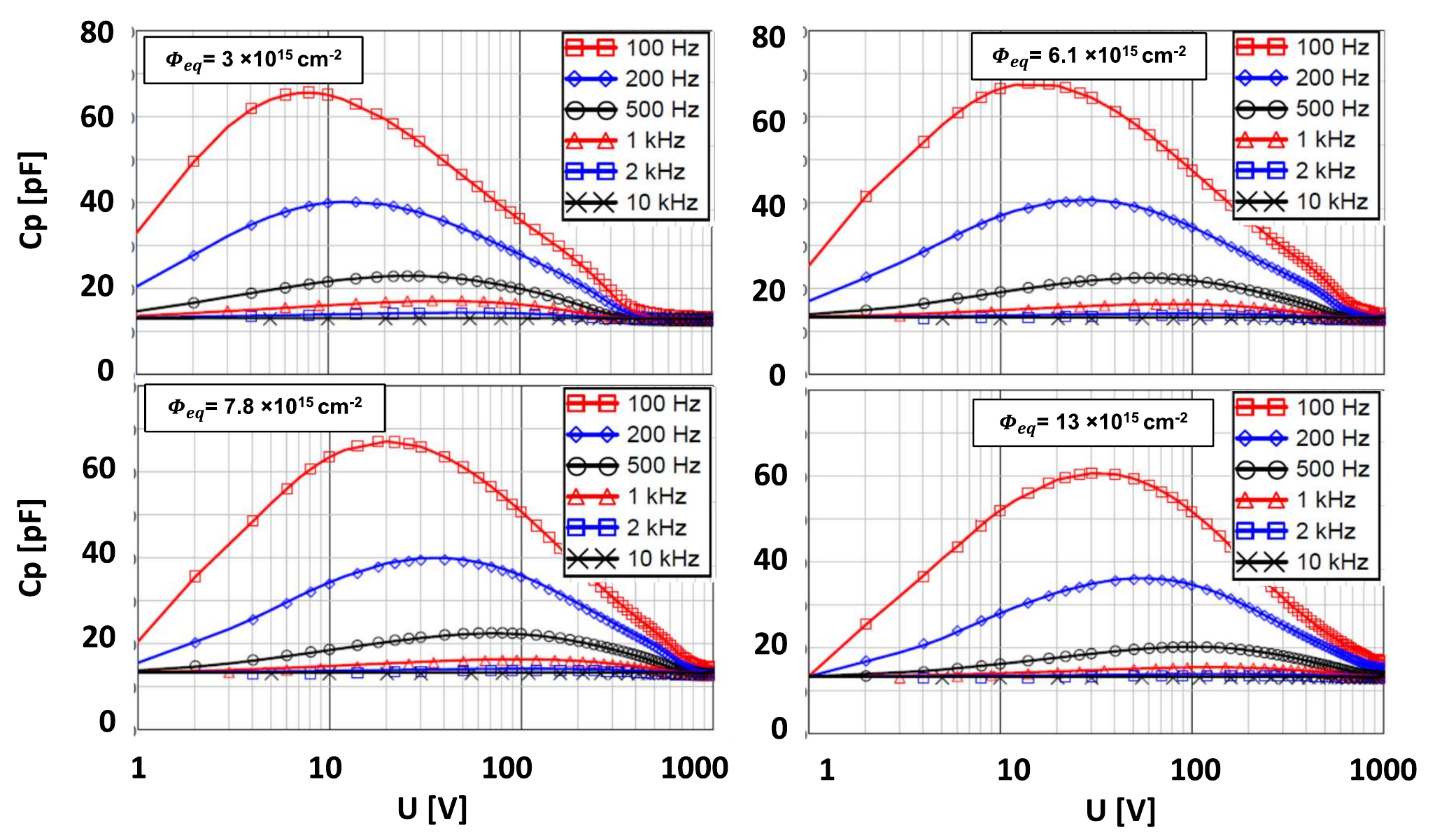}
    \caption{Voltage dependence of $\mathit{Cp}$ for selected frequencies at $T = - 30^{\,\circ}$C.
     The symbols are the experimental data, and the lines the results of the fits. }
  \label{fig:CpU-m30}
 \end{figure}

 Figs.\,\ref{fig:CpU-m20} and \ref{fig:CpU-m30} present $\mathit{Cp}$ for voltages between 1\,V and 1000\,V and selected frequencies, for the four irradiations.
 Independent of $\Phi _\mathit{eq}$ and $U$, the geometrical capacitance
 $C_\mathit{geom} = \varepsilon _\mathit{Si} \cdot A / d$
 is reached for $ f > 2$\,kHz.
 At lower frequencies, $\mathit{Cp}$ increases with $U$, reaches a maximum and approaches $C_\mathit{geom}$ at higher voltages.
 The value of $\mathit{Cp}$ at the maximum and the voltage at which the maximum is reached depend on $f$, $T$ and $\Phi _\mathit{eq}$.
 For $ f = 100$\,Hz and $T = - 20^{\,\circ}$C, $\mathit{Cp}^\mathit{max}$ is between 150 and 170\,pF,  reached at 4\,V for  $\Phi _\mathit{eq} = 3 \times 10^{15}$\,cm$^{-2}$ increasing with fluence to 15\,V for $13 \times 10^{15}$\,cm$^{-2}$.
 At $T = - 30^{\,\circ}$C, the corresponding values for $\mathit{Cp}^\mathit{max}$ are 60 to 70\,pF, reached at 8\,V for $\Phi _\mathit{eq} = 3 \times 10^{15}$\,cm$^{-2}$, and at 35\,V for $13 \times 10^{15}$\,cm$^{-2}$.
 These dependencies are completely different from the $C(U)$ dependence of the pad diodes before irradiation, for which, independent of $f$, $\mathit{Cp} \propto 1/\sqrt{U}$ below the full depletion voltage, and $\mathit{Cp} = C_\mathit{geom}$ above.

\begin{figure}[!ht]
   \centering
   \begin{subfigure}[a]{0.5\textwidth}
    \includegraphics[width=\textwidth]{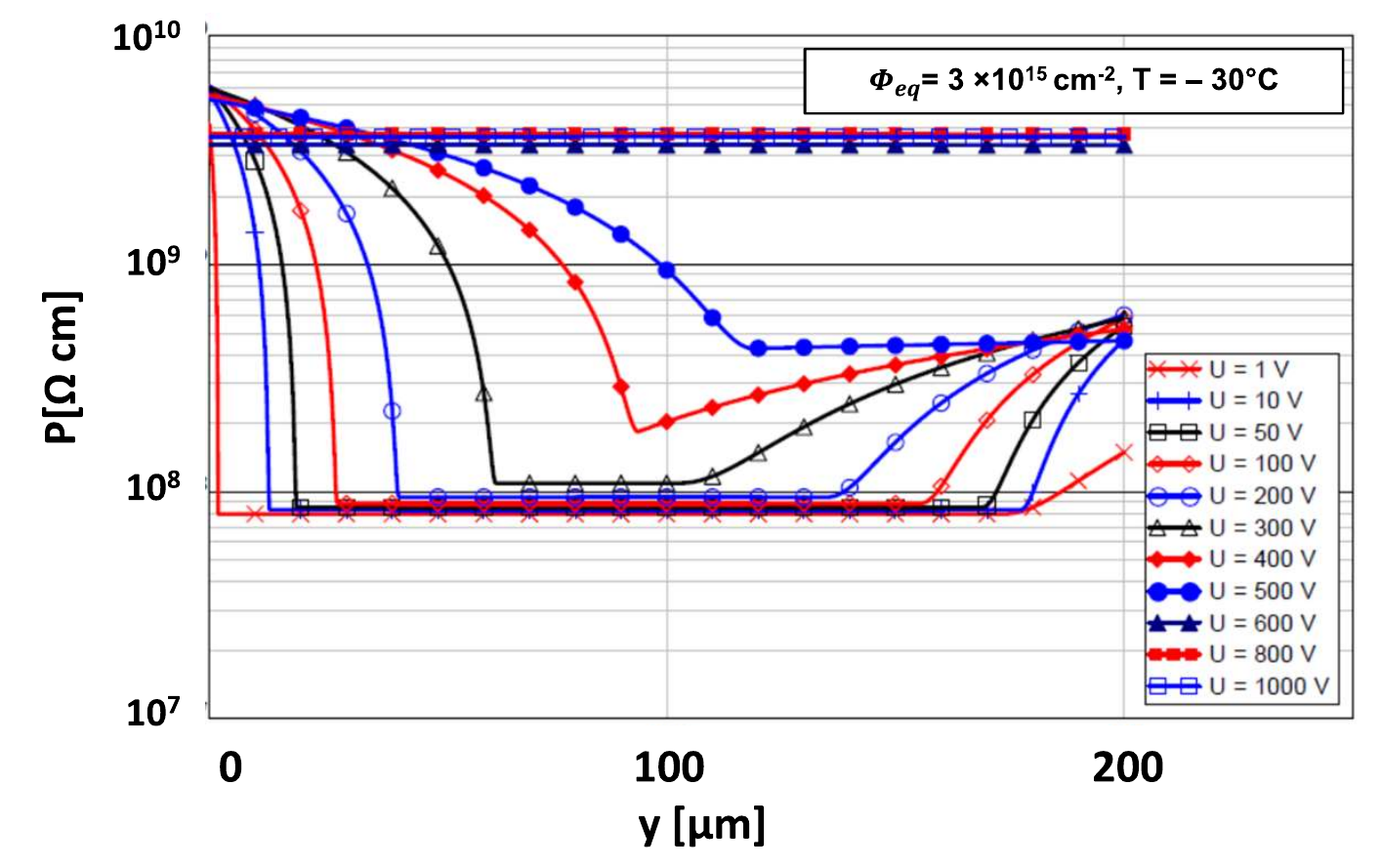}
     \label{fig:rho-m30-3E15}
    \caption{ }
   \end{subfigure}%
    ~
   \begin{subfigure}[a]{0.5\textwidth}
    \includegraphics[width=\textwidth]{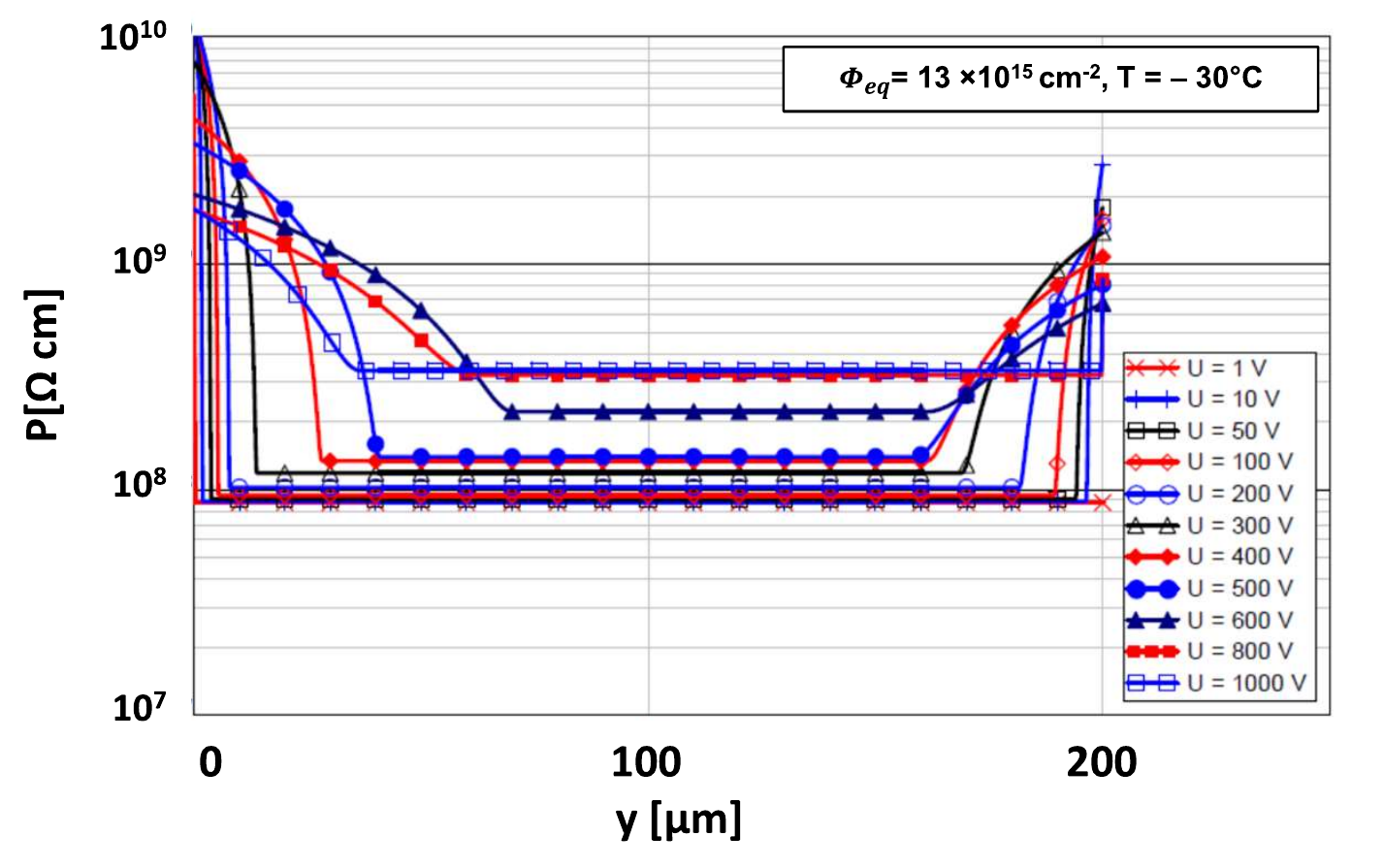}
     \label{fig:rho-m30-13E15}
    \caption{ }
   \end{subfigure}%
   \caption{Results for $\rho (y)$ from the fit to the $Y(f)$ and $I$\,data for selected voltages and $T = - 30^{\,\circ}$C.
   (a) $\Phi _\mathit{eq} = 3 \times 10^{15}$\,cm$^{-2}$, and
   (b) $\Phi _\mathit{eq} = 13 \times 10^{15}$\,cm$^{-2}$.
   As discussed in the text, the fits for $ U > 300$\,V can not be trusted. }
  \label{fig:rho-m30}
 \end{figure}

 Fig.\,\ref{fig:rho-m30} shows the $y$\,dependence of the resistivity, $\rho (y)$, from the fits to the data at $T = - 30^{\,\circ}$C, for $\Phi _\mathit{eq} = 3 \times 10^{15}$ and $13 \times 10^{15}$\,cm$^{-2}$, and $U$ between 1\,V and 1000\,V.
 As long as there is a low-resistivity region, $\rho (y)$ varies by about two orders of magnitude and the result of the fits can be trusted.
 Above a certain voltage, which corresponds to the full-depletion voltage, $U_\mathit{fd}$, the low-$\rho$ region is absent.
 The figure shows that at $- 30^{\,\circ}$C, $U_\mathit{fd} \approx 300$\,V for  $\Phi _\mathit{eq} = 3 \times 10^{15}$\,cm$^{-2}$ increasing to about 500\,V at $13 \times 10^{15}$\,cm$^{-2}$.
 This is expected, because the increase in the density of radiation-induced states results in an increase of the effective doping in the depleted regions.

 Table\,\ref{tab:rho0} shows that for a given temperature $\rho _i$ determined by the fit at $ = 1$\,V is independent of $\Phi _\mathit{eq}$ and approximately agrees with $\rho _\mathit{intr}$ calculated using Eq.\,\ref{equ:rhoi}.
 It should be noted that, mainly due to the uncertainty of $n_i(T)$, the uncertainty of $\rho _\mathit{intr}$ is estimated to be 10\,\%.
 A small linear increase of $\rho _i $ with voltage is observed, which was not expected and so far is not understood.
 However, the slope, which is approximately $7 \times 10^4\,\Omega$/V at $- 20^{\,\circ}$C, and $2 \times 10^5\,\Omega$/V at $- 30^{\,\circ}$C, is small.

 \begin{figure}[!ht]
   \centering
    \includegraphics[width=\textwidth]{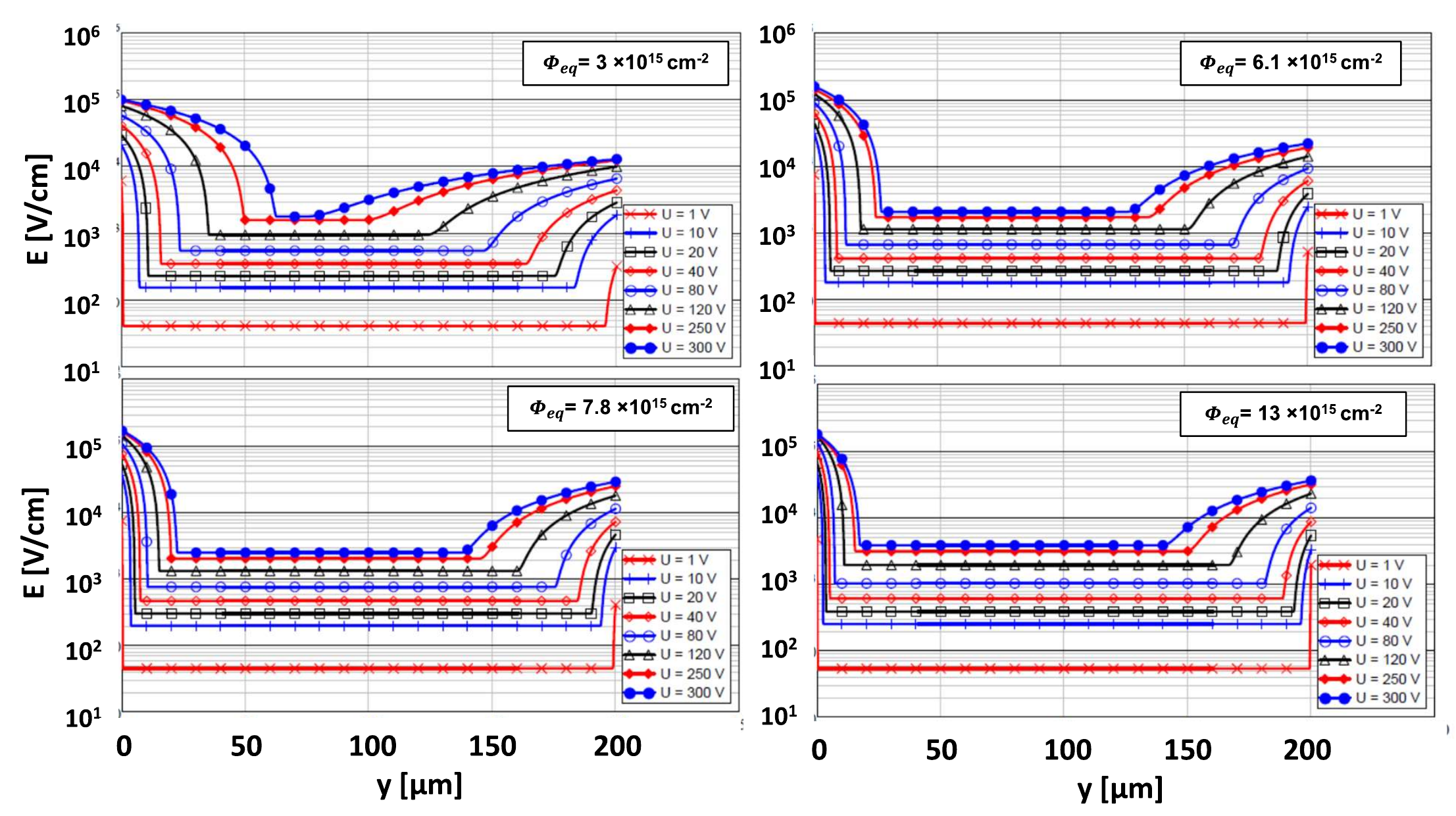}
    \caption{Position dependence of the electric field for different reverse voltages and the four $\Phi _\mathit{eq}$\,values at $T = - 20^{\,\circ}$C.
     The $n^+p$ junction is at $y = 0$.}
  \label{fig:Em20}
 \end{figure}

 \begin{figure}[!ht]
   \centering
    \includegraphics[width=\textwidth]{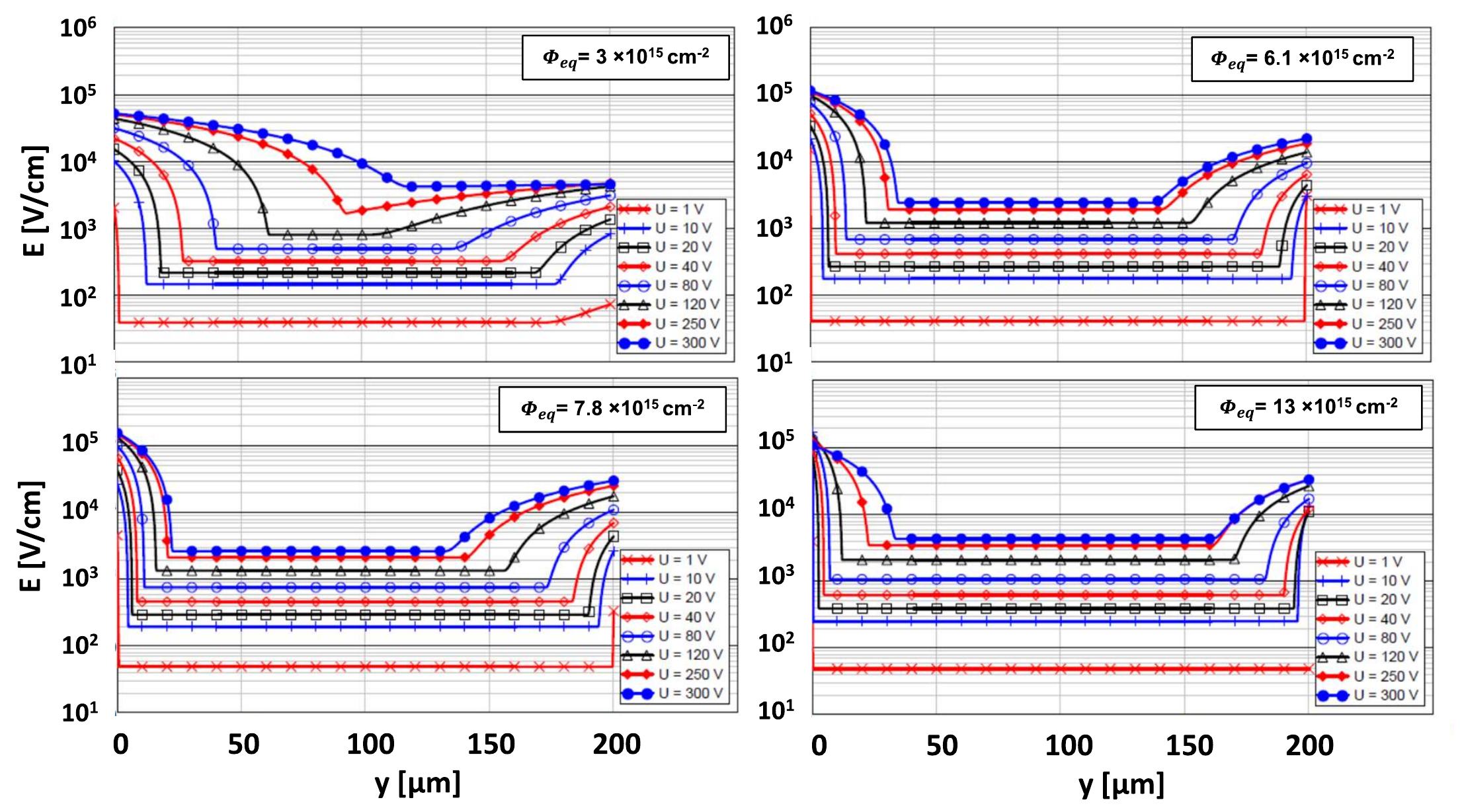}
    \caption{Position dependence of the electric field for different reverse voltages and the four $\Phi _\mathit{eq}$\,values at $T = - 30^{\,\circ}$C.
     The $n^+p$ junction is at $y = 0$.}
  \label{fig:Em30}
 \end{figure}

 Using Eq.\,\ref{equ:rho-E}, the $y$\,dependence of the electric field, $E(y)$, is calculated.
 The results for selected voltages up to 300\,V are shown in Figs.\,\ref{fig:Em20} and \ref{fig:Em30}.
 A high-field region is observed at the $n^+p$\,junction, a low-field region with constant field in the centre, and another high-field region, the \emph{double junction}, towards the rear $p\,p^+$\,junction.
 At low voltages, the low-field region extends over most of the pad diode, and the widths of the high-field regions are just a few $\upmu$m.
 With increasing voltage, the electric fields at both junctions and the widths of both high-field regions increase.
 The field at the $n^+ p$\,junction is about a factor 5 to 10 higher than at the $p\,p^+$\,junction, whereas the width of the high-field region is larger for the latter.
 Similar results are reported in Ref.\,\cite{Klanner:2020}, where the position dependence of the electric field of a $300\,\upmu$m $n^+p$\,strip sensor has been obtained from \emph{edge-TCT} data.
 As discussed before, the voltage at which the central low-field region vanishes corresponds to the full-depletion volte, $U_\mathit{fd}$.


  \subsection{Electric field at high voltages above full depletion}
   \label{sect:HighE}

 In the previous section it has been found that the proposed method for determining $E(y)$ from admittance and current measurements becomes unreliable at higher voltages, when the sensor is fully depleted and the variation of $\rho (y)$ is small.
 This is the case for $U \gtrsim 300$\,V for a fluence $\Phi _\mathit{eq} = 3 \times 10^{15}$\,cm$^{-2}$ and at somewhat higher voltages for higher irradiations.
 However, experiments in a high-luminosity environments aim for fully depleted sensors to achieve high detection efficiencies.
 To illustrate the problems of the method for these conditions, this section presents the results for pad diodes irradiated to $\Phi _\mathit{eq} = 13 \times 10^{15}$\,cm$^{-2}$ for voltages up to 1000\,V.

 \begin{figure}[!ht]
   \centering
   \begin{subfigure}[a]{0.5\textwidth}
    \includegraphics[width=\textwidth]{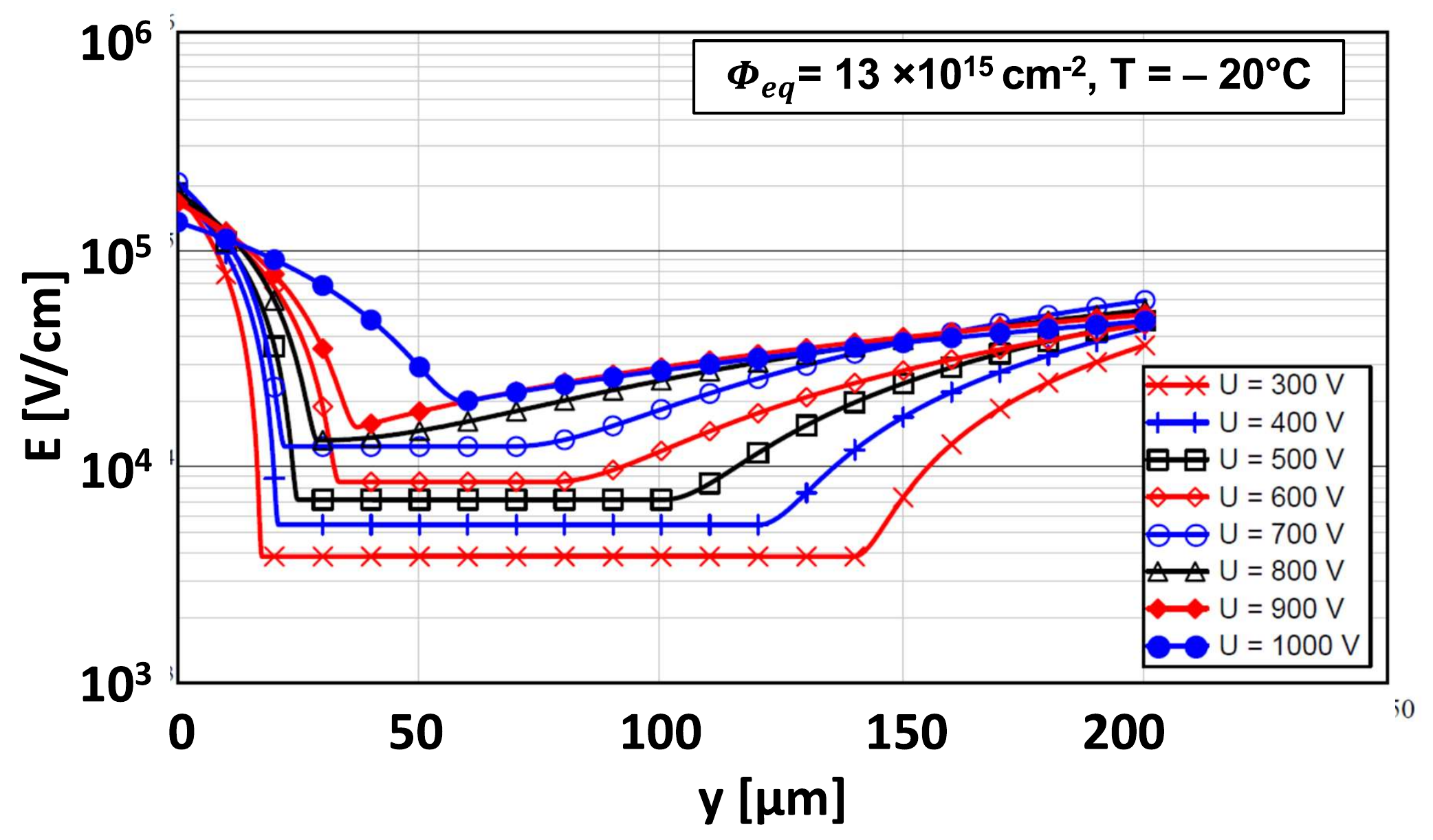}
     \label{fig:E-m20-13E15}
    \caption{ }
   \end{subfigure}%
    ~
   \begin{subfigure}[a]{0.5\textwidth}
    \includegraphics[width=\textwidth]{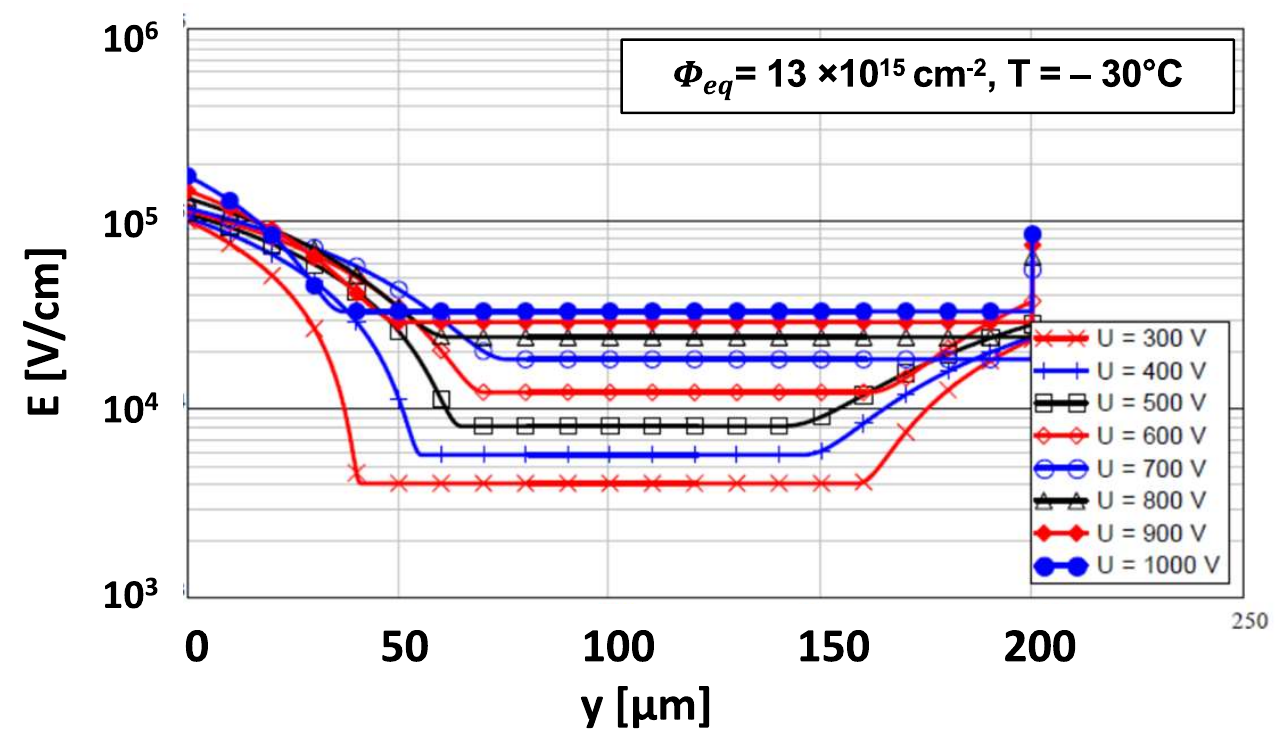}
     \label{fig:E-m30-13E15}
    \caption{ }
   \end{subfigure}%
   \caption{Electric field $E(y)$ for $\Phi _\mathit{eq} = 13 \times 10^{15}$\,cm$^{-2}$ and voltages between 300\,V and 1000\,V at
   (a) $T = - 20^{\,\circ}$C and
   (b) $T = - 30^{\,\circ}$C. }
  \label{fig:E-13E15}
 \end{figure}

 \begin{figure}[!ht]
   \centering
   \begin{subfigure}[a]{0.5\textwidth}
    \includegraphics[width=\textwidth]{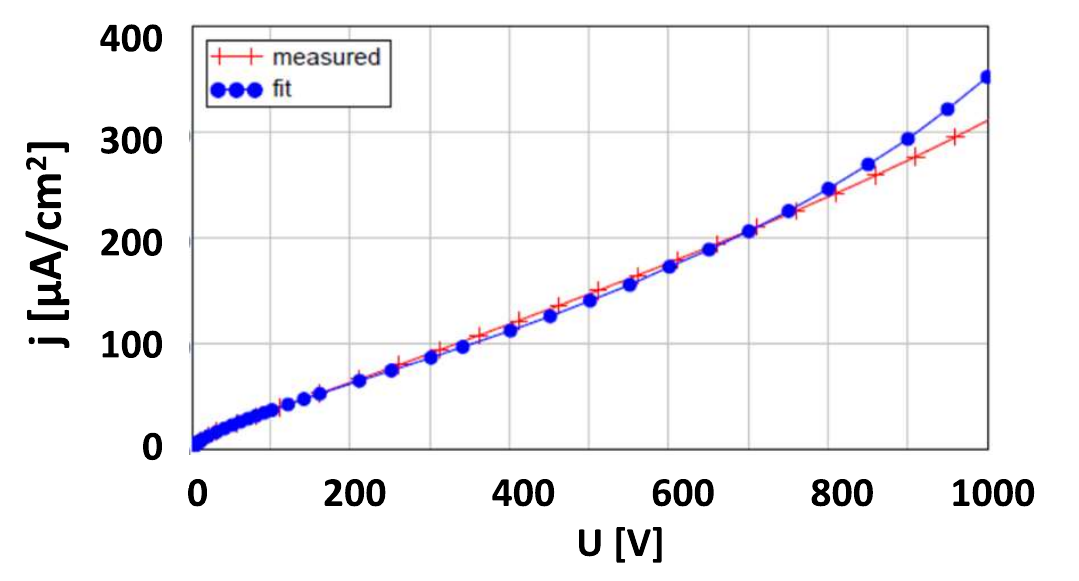}
     \label{fig:IU-m20-13E15}
    \caption{ }
   \end{subfigure}%
    ~
   \begin{subfigure}[a]{0.5\textwidth}
    \includegraphics[width=\textwidth]{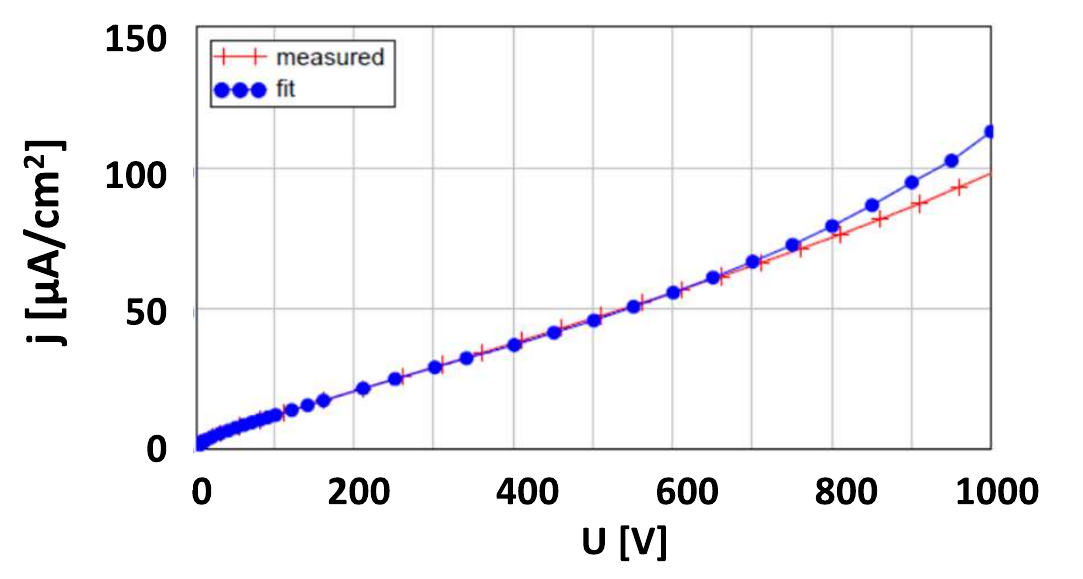}
     \label{fig:IU-m30-13E15}
    \caption{ }
   \end{subfigure}%
   \caption{Comparison of the measured to the fitted current density as a function of voltage for $\Phi _\mathit{eq} = 13 \times 10^{15}$\,cm$^{-2}$  at\,
   (a) $T = - 20^{\,\circ}$C and
   (b) $T = - 30^{\,\circ}$C. }
  \label{fig:IU-13E15}
 \end{figure}

 Fig.\,\ref{fig:E-13E15} shows $E(y)$ for voltages between 300\,V and 1000\,V and $\Phi _\mathit{eq} = 13 \times 10^{15}$\,cm$^{-2}$ at the temperatures of $ - 20^{\,\circ}$C and $ - 30^{\,\circ}$C, and Fig.\,\ref{fig:IU-13E15} compares the measured to the fitted current densities.
 Whereas at lower voltages, the current density is well described by the fit, for voltages above 700\,V the fitted current density is systematically higher than the measured one.
 The electric field, $E(y)$,  in particular at $T = - 30^{\,\circ}$C, shows an unexpected and most probably unphysical behaviour:
 At the $n^+p$\,junction the depletion region shrinks at the highest voltages, and at the $p\,p^+$\,junction a high-field region with a finite extension disappears and is replaced by $\delta $-function at $y = d$.
 In addition, the error matrix of the fit is close to singular, with strong correlations between the parameters.
 Attempts to find a parametrisation which gives physically plausible and at the same time unique results failed.
 It is thus concluded that the method with the parametrisation of Eq.\,\ref{equ:rho_y} does not give reliable results if the ratio of the maximal to the minimal electric field in the diode is less than about a factor 10, which is the case for fully depleted diodes.


  \section{Summary and conclusions}
   \label{sect:Conclusions}

 The topic of this paper is the experimental determination of the electric field in highly-irradiated silicon pad diodes.
 Although the knowledge of the electric field is essential for understanding and simulating the sensor performance and in spite of major efforts over many years, no satisfactory solution has been found so far.
 In this paper the question is investigated to what extent admittance-voltage measurements as a function of frequency together with current measurements can contribute to solving this problem.

 A 1-D model for radiation-damaged pad diodes is proposed which allows calculating the frequency dependence of the admittance $Y = 1/Z$.
 The complex resistance, $Z(\omega )$, is evaluated by summing the position-dependent differential complex resistances $\mathrm{d}Z = (1/\mathrm{d}R + i \cdot \omega \cdot \mathrm{d}C )^{-1}$ over the  diode depth.
 $\mathrm{d}R$ is obtained from the position-dependent resistivity, $\rho$, and $\mathrm{d}C$ is the differential geometric capacitance.
 Possible contributions from the charging and discharging of radiation-induced traps are ignored.
 From the position-dependent $\rho $ and the position-independent current density $j$, the electric field is obtained using Ohm's law in differential form: $E = j \cdot \rho$.
 For the parametrisation of the electric field, a linear increase towards the front and rear faces of the diode and a constant low field in the centre, are used.
 The high-field regions correspond to the depletion regions at the front and rear sides (\emph{double junction}), and the central low-field region to the non-depleted region.

 The model is fitted to admittance data from $200\,\upmu$m thick $n^+ p\, p^+$ pad diodes of 5\,mm $\times $ 5\,mm area irradiated by 24\,GeV/c protons to 1\,MeV neutron equivalent fluences, $\Phi _\mathit{eq}$, of (3, 6.07, 7.75, and 13)\,$\times 10 ^{15}$\,cm$^{-2}$.
 Data have been taken for frequencies between 100\,Hz and 2\,MHz, reverse voltages between 1\,V and 1000\,V, and temperatures of $- 20^{\,\circ}$C and $- 30^{\,\circ}$C.
 The model provides a precise description of the experimental data for voltages up to about 300\,V.
 At both temperatures, the resistivity $\rho$ of the low-field region agrees approximately with the intrinsic resistivity of silicon, as expected for a region where the generation and the recombination of free charge carriers are in equilibrium.
 The electric field distributions obtained from the fit for voltages up to $\approx 300$\,V, agree with expectations and the results from other methods:
 High-field regions at the front and rear side of the diode, with the values and the spatial extensions of the electric field increasing with voltage, and to a lesser extent with $\Phi _\mathit{eq}$.
 The value of the electric field at the $n^+p$ side is found to be a factor 5 to 10 higher than at the $p \, p^+$ side.
 At a given voltage, the extension of the ohmic region increases with  $\Phi _\mathit{eq}$, or in other words, the depletion region decreases.
 It is concluded that the model, with the chosen parametrisation of the position dependence of the electric field, allows a determination of the electric field in partially-depleted, irradiated pad diodes. Contribution from the charging and discharging of the radiation-induced states in the band gap are not required.

 At higher voltages, where the low-field region vanishes and full depletion is reached,
 the variation of the electric field is significantly reduced, the data cannot be fitted with the proposed field parametrisation, and the proposed method does not give reliable results.
 Unfortunately, these are the operating conditions most relevant for the use of silicon detectors at high-luminosity hadron colliders.

 The basic problem of the  method is that both admittance and currents result from integrals over the pad diode and thus their sensitivity to position is limited.
 As a result, a parametrisation of the shape of the electric field is required for obtaining the position dependence of the electric field, and the method only works if the electric field has a large variation over the sensor.
 Nevertheless, the fact that in the ohmic region the intrinsic resistivity is obtained, which differs by a factor three between $-20^{~\circ}$C and $- 30^{~\circ}$C, and that the spatial distribution of the electric field qualitatively agrees with the results from other methods, indicates that the method appears to be valid.
 It should be noted that methods like edge-TCT (Transient-Current-Technique \cite{Kramberger:2010}) or TPA-TCT (Two-Photon-Absorption \cite{Garcia:2017}) also have difficulties in determining the electric fields in thin sensors at high voltages:
 The pulse durations are only 1 to 2\,ns and the signals are small, which makes it essentially impossible to reliably extract the electric field from the initial shape of the transients\,\cite{Klanner:2020}.
 For further progress, position-sensitive methods could be combined with the method proposed in this paper or completely different methods like the one proposed in Ref.\,\cite{Klanner:2021} should be developed.

 The answer to the question asked in the title:
 \emph{Can the electric field in radiation-damaged silicon pad diodes be obtained from admittance and current measurements?}
 is a partial \emph{yes} only.
 The main results of the investigation are:
 Precise measurements of the admittance and current provide valuable information on the electric field distribution in radiation-damaged silicon sensors.
 The frequency dependence of the admittance can be described without considering the charging and discharging of the radiation-induced levels in the silicon band gap.
 The electric field can be determined as long as the sensor is partially depleted, but the proposed method fails for fully depleted sensors.

 It is concluded that the determination of the electric fields in highly-irradiated silicon sensors for the conditions most relevant for experiments in high-radiation environments remains a major challenge.

  \begin{appendices}

  \section{Resistivity, electric field, generation- and recombination-rates}
   \label{sect:Appendix}

 In this Appendix the systematic error of the $E$\,determination using Eq.\,\ref{equ:rho-E}, which ignores the diffusion contribution to the current density, is estimated.
 The one-dimensional steady-state equations for the electron- and hole-current densities, $j_e$ and $j_h$, are\,(\cite{Sze:2007})
 \begin{equation} \label{eq:jejh}
   j_e = \mu _e \cdot (q_0 \cdot n_e \cdot E + k_B T \cdot \mathrm{d}n_e /\mathrm {d}x)
    \hspace{3mm} \mathrm{and} \hspace{3mm}
   j_h = \mu _h \cdot (q_0 \cdot n_h \cdot E - k_B T \cdot \mathrm{d}n_h /\mathrm {d}x),
 \end{equation}
 and the continuity equations
 \begin{equation} \label{equ:j_cont}
   q_0 \cdot \mathit{GR} + \mathrm{d}j_e /\mathrm {d}x = 0
    \hspace{3mm} \mathrm{and} \hspace{3mm}
   q_0 \cdot \mathit{GR} - \mathrm{d}j_h /\mathrm {d}x = 0\,,
 \end{equation}
 with $\mathit{GR}(x) = \mathrm{d}n_e(x) /\mathrm {d}t = \mathrm{d}n_h(x) /\mathrm {d}t$, the position-dependent difference of volume generation and recombination rate of electron-hole pairs including charge-carrier multiplication at high electric fields.
 Inserting
 \begin{equation} \label{equ:dndx}
   \mathrm{d}n_e /\mathrm {d}x = (\mathrm{d}n_e /\mathrm {d}t)/(\mathrm{d}x /\mathrm {d}t) = - \mathit{GR}/(\mu _e \cdot E)
    \hspace{3mm} \mathrm{and} \hspace{3mm}
   \mathrm{d}n_h /\mathrm {d}x = (\mathrm{d}n_h /\mathrm {d}t)/(\mathrm{d}x /\mathrm {d}t) = + \mathit{GR}/(\mu _h \cdot E)
 \end{equation}
 into Eq.\,\ref{eq:jejh}, and using $1/ \rho = q_0 \cdot (\mu _e \cdot n_e + \mu _h \cdot n_h) $ gives the total current density, $j$, which in steady-state conditions does not depend on $x$
 \begin{equation} \label{equ:j_cont}
   j = j_e (x) + j_h (x) = E(x)/\rho(E) - 2\,k_B T \cdot \mathit{GR}(x) / E(x).
 \end{equation}
 The solution of this second order equation in $E$, which in the limit of $GR = 0$ agrees with Ohm's law is
 \begin{equation} \label{equ:Ejrho}
   E(x) = \frac{j \cdot \rho (x)}{2} \cdot \left(1 + \sqrt{1 + \frac{8 \cdot k_B T \cdot \mathit{GR(x) } } {j^2 \cdot \rho (x)}} \right)
   \approx j \cdot \rho (x) + \frac{2\,k_B T \cdot \mathit{GR} (x) }{j}.
 \end{equation}
  The approximation for a small correction to $E = j \cdot \rho $ is given on the right.
 For the calculation of $E(x)$, in addition to $\rho (x)$, $\mathit{GR} (x)$, has to be known.
 An  approximation is:
 $\mathit{GR} = 0 $ in the non-depleted region, where generation and recombination are the same, and $\mathit{GR} = j / (q_0 \cdot w) $ in the depleted regions of total depth $w$.
 The latter assumes no charge multiplication and a position-independent \emph{GR} in the depletion regions.
 The final results is a correction term which only depends on $w$ and $T$.
 \begin{equation} \label{equ:E_approx}
   E(x) \approx \rho (x) \cdot j + \frac{2\,k_B T }{q_0 \cdot w}.
 \end{equation}
 Typical values for the correction term at $- 20^{\,\circ}$C are 2.2\,V/cm for $w = 200 \,\upmu$m, and 45\,V/cm for $w = 10 \,\upmu$m, which justifies using Eq.\,\ref{equ:rho-E}.
 High-field effects, like multiplication cause a strong position dependence of \emph{GR} and larger values of the correction term are expected.
 However, they will occur in the high-field regions with minor influence on the relative value of $E$.

  \end{appendices}




  \section*{Bibliography}
   \label{sect:Bibliography}


\begin{thebibliography}{9}

  \bibitem{Schroder:2006}
   D.~K.~Schroder,
    \emph{Semiconductor Material and Device Characterization},
     IEEE Press, A Wiley-Interscience Publication, Third Edition, 2006.

  \bibitem{Schwandt:2020}
   J.~Schwandt and R.~Klanner,
   \emph{On the weighting field of irradiated silicon detectors},
    Nucl. Instr. \& Methods A~951 (2020) 162982.

  \bibitem{Wilson:1968}
   D.~K.~Wilson,
    \emph{Capacitance Recovery in Neutron-Irradiated Silicon Diodes by Majority and Minority Carrier Trapping},
    IEEE Trans. Nucl. Sci. NS--15 (1968) 77--83.

  \bibitem{Tokuda:1977}
   Y.~Tokuda and A.~Usami,
    \emph{Admittance studies of neutron-irradiated slicon $p^+n$ diodes},
    Journal of Applied Physics 48 (1977) 1668--1672.

  \bibitem{Losee:1972}
   D.~L.~Losee, \emph{Admittance spectroscopy of deep impurity levels: ZnTe Schottky barriers}, Journal of Applied Physics 21 (1972) 54--56.

  \bibitem{Losee:1975}
   D.~L.~Losee,
   \emph{Admittance spectroscopy of impurity levels in Schottky barriers},
    Journal of Applied Physics 46 (1975) 2204--2214.

  \bibitem{Vincent:1975}
   G.~L.~Vincent, D.~Bois and P.~Pinard,
    \emph{Conductance and capacitance ctudies in GaP Schottky barriers},
    Journal of Applied Physics 46 (1975) 5173--5178.

  \bibitem{Li:1990}
   Z.~Li and H.~Kraner,
    \emph{Studies of frequency dependent $C-V$ characteristics of neutron irradiated $p^+n$ silicon detectors},
     Brookhaven Natl. Laboratory Internal Report, BNL-44648 (1990).

  \bibitem{Croitoru:1997}
   N.~Croitoru, et al.,
    \emph{Dependence of the a.c. conductance of neutron irradiated silicon detectors on frequency and temperature},
     Nucl. Instr. \& Methods A~386 (1997) 156--161.

  \bibitem{Hamamatsu}
   Hamamatsu Photonics K.K., \url{www.hamamatsu.com}.

  \bibitem{CERN_Irr}
   CERN Irradiation facility,
    \url{https://ps-irrad.web.cern.ch/ps-irrad/}.

  \bibitem{Eremin:1995}
   V.~Eremin, Z.~Li and I.~Ilyashenko,
    \emph{Trapping induced $N_\mathit{eff}$ and electrical field transformation at different temperatures in neutron irradiated high resistivity silicon detectors,}
   Nucl. Instr. \& Methods A~360 (1995) 458--462.

  \bibitem{Klanner:2020}
   R.~Klanner, et al.,
   \emph{Determination of the electric field in highly-irradiated silicon sensors using edge-TCT measurements},
    Nucl. Instr. \& Methods A~951 (2020) 162987.

  \bibitem{Scharf:2018}
   C.\,Scharf,
    \emph{Radiation damage of highly irradiated silicon sensors},
     PhD thesis, University of Hamburg 2018, doi:10.3204/PUBDB-2018-03707.

  \bibitem{Lombardi:1988}
   C.~Lombardi, et al.,
    \emph{A physically based mobility model for numerical simulation of nonplanar devices},
     IEEE Trans. on Computer-Aided Design, vol.~77 (1988) 1164--1171.

  \bibitem{Green:1990}
   M.~Green,
    \emph{Intrinsic concentration, effective densities of states, and effective mass in silicon,}
         Journal of Applied Physics 67 (1990) 2944--2954.

 \bibitem{Kramberger:2010}
 G. Kramberger, et al.,
  \emph{Investigation of Irradiated Silicon Detectors by Edge-TCT},
  IEEE Trans. Nucl. Sci. 57 (2010) 2294--2302.

  \bibitem{Garcia:2017}
   M. Fernandez Garcia, et al.,
   \emph{On the determination of the substrate effective doping concentration of irradiated HV-CMOS sensors using an edge-TCT technique based on the Two-Photon-Absorption process,}
    2017 JINST 12 C01038.

  \bibitem{Klanner:2021}
   R. Klanner and A. Vauth,
   \emph{Electro-optical imaging of electric fields in silicon sensors},
    Nucl. Instr. \& Methods A~995 (2021) 165082.

  \bibitem{Sze:2007}
   S.M.\,Sze, \emph{Physics of  Semiconductor Devices},
    A Wiley-Interscience Publication, John Wiley \& Sons, Third Edition, 2007.





 \end{thebibliography}
\end{document}